\documentclass[letterpaper,preprint, aps,superscriptaddress,floatfix,showpacs,showkeys]{revtex4-1}
\usepackage{hyperref}
\usepackage{amsmath}
\usepackage{graphicx}
\usepackage{caption}
\usepackage{subcaption}
\usepackage{natbib}

\begin{document}

\title{Covariance and correlation estimators in bipartite complex systems with a double heterogeneity}
\date{\today}

\author{Elena Puccio}
\email{elena.puccio@unipa.it}
\affiliation{Dipartimento di Fisica e Chimica, Universit\`a di Palermo, Viale delle Scienze, 90128 Palermo, Italy}
\author{Jyrki Piilo}
\affiliation{Turku Centre for Quantum Physics, Department of Physics and Astronomy, University of Turku, FI-20014 Turun yliopisto, Finland}
\author{Michele Tumminello}
\affiliation{Dipartimento di Scienze Economiche, Aziendali e Statistiche, Universit\`a di Palermo, Viale delle Scienze, 90128 Palermo, Italy}

\begin{abstract}
We present a weighted estimator of the covariance and correlation in bipartite complex systems with a double layer of heterogeneity. The advantage provided by the weighted estimators lies in the fact that the unweighted sample covariance and correlation can be shown to possess a bias. Indeed, such a bias affects real bipartite systems, and, for example, we report its effects on two empirical systems, one social and the other biological. On the contrary, our newly proposed weighted estimators remove the bias and are better suited to describe such systems.
\end{abstract}



\maketitle

\section*{Introduction}

Bipartite systems consist of two sets of elements in which elements of one set directly relate to elements of the other set only. Often these systems are described as networks. Complete information about bipartite systems can usually be incorporated in bipartite networks, however, many studies use the bipartite structure of the system only to set relationships between the elements of one of the two sets. For instance, the scientific collaboration network in \cite{Newman_1}, \cite{Newman_2} can be seen as the projection of the bipartite system of authors and papers, where co-authored papers are only used to set a relationship between any pair of authors.

Bipartite networks and their projections are widely used to study complex systems such as mobile communication \cite{Onnela2007a,Onnela2007b}, criminal activity \cite{Tumminello2013}, interbank credit markets~\cite{Iori2008,Hatzopoulos2015}, investors activity~\cite{Tumminello2014}, and recommendation systems for users and objects~\cite{PR_2012,Fiasconaro2015}. A common feature of complex bipartite systems is heterogeneity, which typically characterizes both sides of the system and makes the statistical analysis of the various properties a challenging task. This is apparent in the scientific collaboration network, where there is heterogeneity of authors in terms of the number of papers they authored, and heterogeneity of papers in terms of the number of co-authors. Indeed, Newman -- to account for such heterogeneity -- assigned a weight to the link between two coauthors that takes into account information about the number of authors of each paper they have in common~\cite{Newman_2}.

However, when one is interested in covariance and correlation coefficients, we show that even Newman's solution is not sufficient to account for the double heterogeneity present in complex bipartite systems. In general, the presence of such heterogeneity of degree may induce a bias in covariance and correlation coefficient estimates, which, in turn, would make the task of discriminating information from noise in covariance/correlation matrices even more impervious \cite{Laloux_1999}, \cite{Plerou_1999}, \cite{MacMahon_2015}.

To remove such a bias from covariance and correlation coefficients we introduce weighted estimatorsthat take into account, at once, the heterogeneity  on both sides of a bipartite network. Moreover, we also quantify the improvement of the new estimators compared to unweighted ones and demonstrate the power of the introduced methodology with applications to two real social and biological datasets. From a conceptual point of view, the newly proposed estimators are such that the covariance/correlation between any two given elements in the system depends on all the others, in such a way that adding or removing even a single element influences the value of the estimator. To prove the stability of the weighted estimators against such a change in the system, we ran a robustness analysis and show that the proposed estimators are rather robust to changes in the system composition up to 30\%. 

The paper is structured in the following way. Section \ref{sec:I} discusses the problem of a bias in the sample covariance and correlation of bipartite systems and in Section \ref{sec:II} we propose a model of the rewiring process which demonstrates that the expected value of the covariance is different from zero. In Section \ref{sec:III} we define the new weighted covariance estimator in the multivariate case and show that its expected value is indeed null, and in Section \ref{sec:IV} we focus on the weighted correlation coefficient and show the improvement it offers over the unweighted one. Section \ref{sec:V} displays the results of employing the weighted against the unweighted estimators in two empirical datasets.

\section{\label{sec:I} Sample covariance and correlation in bipartite systems}

\begin{figure}[!htb]
	\includegraphics[width=0.9\textwidth]{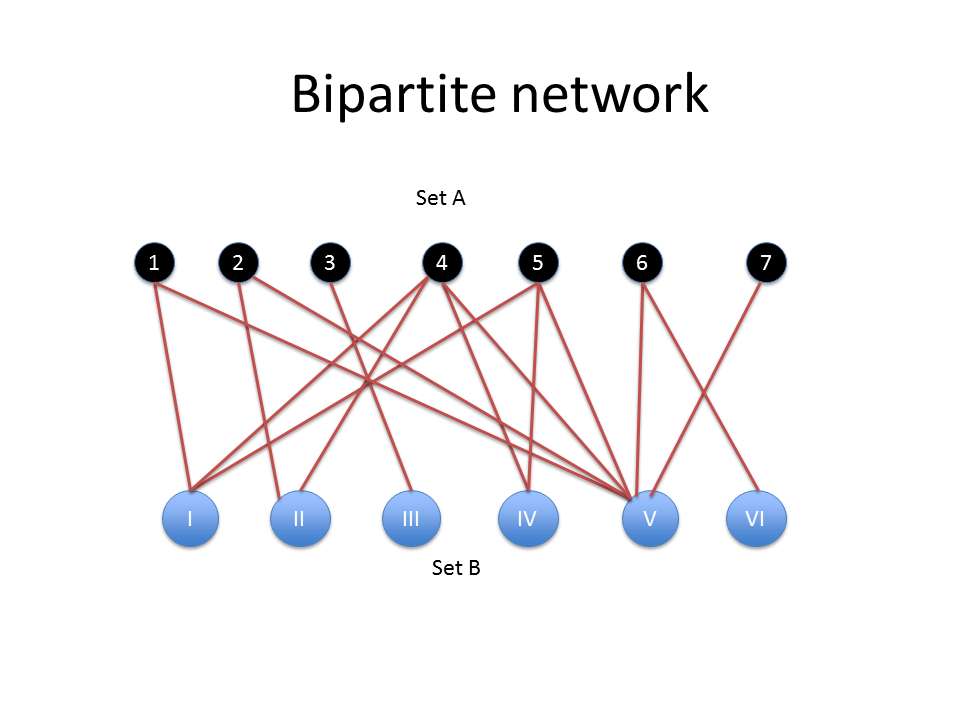}
    \caption{Schematic representation of a bipartite network with $N$ nodes in set $A$ (black), e. g., authors, and $T$ nodes in set $B$ (blue), e. g., papers. Links are only possible between the two sets and are shown in red. A projected network of nodes in set $A$ is obtained by linking any two nodes in $A$ that share one or more connections to nodes in set $B$ of the bipartite network.}
    \label{fig:bip}
\end{figure}

In bipartite networks elements can be divided in two disjoint, independent sets, such that only links between the two sets are allowed, see Fig. \ref{fig:bip}.

Let's suppose we measure the sample covariance between two elements $i$ and $j$ in set $A$ of a bipartite system, as the scalar product between the binary vectors $\mathbf{{v}_i}$ and $\mathbf{{v}_j}$. A component ${v}_{i,h}$ (${v}_{j,h}$), with $h \in [1,...,T]$, of vector $\mathbf{{v}_i}$ ($\mathbf{{v}_j}$) is equal to 1 if element $i$ ($j$) is linked to node $h$ in set $B$, and 0 otherwise. Therefore, the sample covariance estimator between two binary vectors can be written as \cite{Fiasconaro2015}:
\begin{equation}
\hat{\sigma}_{ij}=\frac{1}{T}\left(\mathbf{{v}_i} \cdot \mathbf{{v}_j}\right)-\frac{1}{T^2} \, \left(\sum_{h=1}^T v_{i,h} \right) \left(\sum_{h=1}^T v_{j,h} \right)= \frac{1}{T} \left(\hat{n}_{ij}-\frac{K_i \, K_j}{T} \right),
\label{eq:cov}
\end{equation}
the hat is henceforth used to denote an estimator, as opposed to its theoretical counterpart. In Eq. (\ref{eq:cov}) $\hat{n}_{ij}$ is the observed number of links in common between the pair of elements $i$ and $j$, of degree $K_i=\sum_{h=1}^T v_{i,h}$ and $K_j=\sum_{h=1}^T v_{j,h}$. Degrees are parameters which are kept fixed throughout. For example, looking at Fig. \ref{fig:bip}, we have for the pair of nodes $4$ and $5$ in set $A$, of degree, respectively,  $K_4=4$ and $K_5=3$, binary vectors $\mathbf{{v}_4}=\{ 1,1,0,1,1,0 \}$ and $\mathbf{{v}_5}=\{ 1,0,0,1,1,0  \}$, number of common links $n_{45}=3$, a covariance of $\hat{\sigma}_{45}=3-3 \cdot 4/6=1$.

From Eq. (\ref{eq:cov}), the sample correlation coefficient estimator between two binary vectors becomes:
\begin{equation}
\hat{\rho}_{ij}= \frac{\hat{\sigma}_{ij}}{\hat{\sigma}_i \, \hat{\sigma}_j}=\frac{\hat{n}_{ij}-\frac{K_i \, K_j}{T}}{\sqrt{K_i  \left(1-\frac{K_i}{T} \right)  K_j  \left(1-\frac{K_j}{T} \right)}},
\label{eq:Pearson_c}
\end{equation}
where $\hat{\sigma}_i$ and $\hat{\sigma}_j$ are standard deviation estimators of vector $\mathbf{{v}_i}$ and $\mathbf{{v}_j}$,
\begin{equation*}
\hat{\sigma}_i=\sqrt{\frac{K_i}{T} \left(1- \frac{K_i}{T}\right)}, \ \ \ \ \ \ \ \hat{\sigma}_j=\sqrt{\frac{K_j}{T} \left(1- \frac{K_j}{T}\right)}.
\end{equation*}

As it often happens when dealing with correlations, the evaluation of noise plays a crucial role. For what concerns bipartite systems, the most widely used procedure to deal with noise is to use a randomized network as a null model, by performing a rewiring of the full bipartite system \cite{Vespignani_NatPhys_2006}. Basically, one step in the rewiring procedure consists in randomly sampling a pair of links in the bipartite network, involving two nodes on each side, and a swap of the target nodes of the link in set B, if the latter newly formed links are not already present in the system. For example, from Fig. \ref{fig:bip}, one randomly selects the pair of links $4-II$ and $6-IV$ and swaps the target nodes in set B to obtain two new links $4-IV$ and $6-II$, since neither $4$ nor $6$ were already linked, respectively, to $IV$ and $II$. In order to randomize the network, one needs to perform a great number of swaps, stopping when the overlapping between the original and rewired networks, evaluated with an appropriate measure, stabilizes around a minimum value (see Section \ref{sec:V} for details).

Unfortunately, we found out that even rewiring the bipartite network, cannot get rid of all the noise present in the covariance and correlation matrices. The residual noise still present in the rewired network appears to depend on both sets' degree distributions, that is, on the intrinsic double heterogeneity of the system. Thus, the sample covariance and correlation estimators appear biased in such systems, and the bias won't be uniform.

\section{\label{sec:II} Expected value of the covariance and correlation under a biased urn model}

Here, we propose a model which approximately describes the statistical properties of the outcome of a random rewiring procedure. The model we propose is a simplification of the problem which, nonetheless, allows us to exactly preserve the degree distribution on one side of the bipartite network, and to keep the degree distribution on average on the other side. The underlying idea is to model the random rewiring as a sampling from a biased urn, followed by a sampling from an unbiased urn, both without replacement (to preserve degrees). 

Our aim is to show the origin of the bias in the covariance and correlation coefficient in Eqs. (\ref{eq:cov}) and (\ref{eq:Pearson_c}) of the randomized network, by calculating their expected values and showing that they are different from zero.

To show the presence of a bias we describe a simplified situation, where nodes in set $B$ only have either a high degree, which we'll formalize as a heavy weight $w_2$, or a low degree $w_1$ (a "light" weight). If we now look at how random links form between a node $i$ in set $A$ and a number $K_i$ of nodes in set $B$, such a process can be modeled as a sampling of exactly $K_i$ marbles (node's $i$ degree), from the total of $T$ marbles in set $B$. The crucial hypothesis is that we assume that marbles have two different probabilities of being selected. Specifically, $m$ marbles have a probability to be sampled proportional to weight $w_2$ (heavy), whereas the remaining $T-m$ marbles have a probability to be sampled proportional to $w_1$ (light), and we define the weight ratio as $w=w_2/w_1>1$. The weight models the heterogeneity in set $B$. We'll focus on Eq. (\ref{eq:cov}), and show that the expected value of $\sigma_{ij}$ is, in general, different from zero, if $w>1$. 

In this model, each node $i$ in set $A$ samples a total of $K_i$ marbles, of which $k_i^w$ are heavy and the remaining $K_i- k_i^w$ are light. In a biased urn problem without replacement, a single variable $w$ is sufficient to describe the system, with the stochastic variable $k_i^w \in [\max(0, K_i-T+m), \min(K_i, m)]$ following the Wallenius non-central hypergeometric distribution \cite{Wallenius}.

If all marbles are distinguishable, for example labeled, we now ask ourselves what would be the intersection $n_{ij}$ between the marbles sampled by two different nodes, $i$ and $j$, in $A$. The expected number of sampled objects $\mathbf{E}[n_{ij}|k_i^w, k_j^w]$ in common between $i$ and $j$ will be the sum of the expected number of heavy marbles in common, $n_{ij}^w$, and the expected number of light ones in common, $n_{ij}^1$,
\begin{equation*}
\mathbf{E}[{n}_{ij}|k_i^w, k_j^w]=\mathbf{E}[n_{ij}^w|k_i^w, k_j^w]+\mathbf{E}[n_{ij}^1|k_i^w, k_j^w].
\end{equation*}

The underlying probability distribution, since each weight-group is now homogeneous, is the Hypergeometric distribution. Specifically, the probability that both nodes sampled exactly $n_{ij}^w$ heavy marbles in common out of the $m$ available ones is given by $P(n_{ij}^w;k_i^w, k_j^w,m)$, and, for what concerns the light marbles in common, $P(n_{ij}^1; K_i-k_i^w,K_j-k_j^w,T-m)$. Since the sampling processes are indipendent, variables $n_{ij}^w$ and $n_{ij}^1$ are independent as well, so that the joint probability distribution is just the product of the previous two.
The expected numbers of common heavy and light marbles can be easily calculated,
\begin{equation*}
\mathbf{E}[n_{ij}^w|k_i^w, k_j^w]=\frac{k_i^w \, k_j^w}{m} \ \ \ \text{and} \ \ \ \mathbf{E}[n_{ij}^1|k_i^w, k_j^w] = \frac{( K_i-k_i^w)( K_j-k_j^w)}{T-m},
\end{equation*}
thus the expected number of marbles in common between $i$ and $j$ turns out to be:
\begin{equation}
\mathbf{E}[{n}_{ij}]=\sum_{ k_i^w, k_j^w} \left(\mathbf{E}[n_{ij}^w| k_i^w, k_j^w]+\mathbf{E}[n_{ij}^1| k_i^w, k_j^w] \right) \, W( k_i^w) \, W( k_j^w)=   \frac{\mu_i \, \mu_j }{m}  + \frac{(K_i- \mu_i ) (K_j- \mu_j)}{T-m},
\label{eq:exp_nij}
\end{equation}
where $\mu_i$ ($\mu_j$) is the expected value of $k_i^w$ ($k_j^w$) calculated with the Wallenius distribution PMF $W(k_i^w)$ ($W(k_j^w)$).

Unfortunately, no exact formula for the mean of the Wallenius distribution is known \cite{Wallenius}, however, the approximate solution of the following equation is reasonably accurate \cite{Manly_1974}:
\begin{equation}
\frac{\mu_i}{m}+ \left( 1-\frac{K_i-\mu_i}{T-m} \right) ^w =1.
\label{eq:w_mean}
\end{equation}

\begin{figure}[!htb]
  \includegraphics[width=0.45\linewidth]{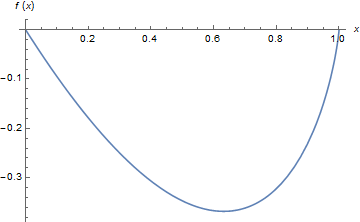}
  \includegraphics[width=0.45\linewidth]{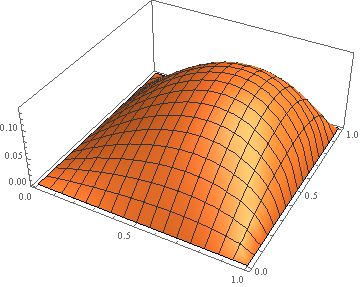}
\caption{Left panel: plot of $f(x)=(1-x) \ln(1-x)$ for $x \in [0,1]$, the function is strictly negative and displays a minimum in $x_m=1-1/e\simeq 0.632$. 
Right panel: 3D plot of $f(x,y)=(1-x) \ln(1-x)\cdot (1-y) \ln(1-y)$ for $ x,y \in [0,1]$, the function is strictly positive and shows a maximum in $\{x_M,y_M\}=\{1-1/e,1-1/e\}$.}
\label{fig:fx}
\end{figure}

Finally, by calculating the Taylor series up to second order of $\mathbf{E}[{n}_{ij}]$ in Eq. (\ref{eq:exp_nij}) near $w=1$ and due to the linearity of operator $\mathbf{E}[]$, the expected value of the covariance can be approximated by:
\begin{eqnarray}
\mathbf{E}[\sigma_{ij}]&=& \frac{\mathbf{E}[{n}_{ij}]}{T} -\frac{K_i \, K_j}{T^2} \simeq \nonumber \\
& \simeq &  \frac{m(T-m)}{T^2} \left[ \left( 1-\frac{K_i}{T} \right) \ln \left( 1-\frac{K_i}{T} \right) \right] \left[ \left( 1-\frac{K_j}{T} \right) \ln \left( 1-\frac{K_j}{T} \right) \right] (w-1)^2.
\label{eq:approx_sij}
\end{eqnarray}
For a graphical representation of the dependency of $\mathbf{E}[\sigma_{ij}]$ on $K_i,K_j$ see Fig.~\ref{fig:fx}.

The expected value of the correlation coefficient in Eq. (\ref{eq:Pearson_c}) can be calculated from Eq. (\ref{eq:approx_sij}) dividing by the standard deviations, which depend only on fixed parameters:
\begin{equation}
\mathbf{E}[\rho_{ij}]  \simeq \frac{m(T-m)}{T\sqrt{K_i \! \left(1-\frac{K_i}{T} \right) \! K_j  \left(1-\frac{K_j}{T} \right)}} \!  \left( 1-\frac{K_i}{T} \right) \! \ln \! \left( 1-\frac{K_i}{T} \right) \! \left( 1-\frac{K_j}{T} \right) \! \ln \! \left( 1-\frac{K_j}{T} \right) \! (w-1)^2.
\label{eq:approx_rhoij}
\end{equation}

From Eq. (\ref{eq:approx_sij}) and Eq. (\ref{eq:approx_rhoij})  it's easy to see how the expected value of both the covariance and the correlation coefficient depends on $i$'s and $j$'s degrees, $K_i$ and $K_j$, as well as on $w$, which is the ratio of $w_2$ to $w_1$ (here representing the heterogeneity of the other set, $B$, in the bipartite system). Thus, we've shown there exists a bias due to the interplay between both sets' heterogeneity in a bipartite system. Now, we'd like to get rid of this bias, by defining new estimators of the covariance and correlation coefficient, whose expected value is zero.

\section{\label{sec:III} Multivariate weighted covariance estimator}

In the most general case, we're dealing with $n<T$ groups, each containing $\mathbf{m}=\{m_1,m_2,...,m_n\}$ marbles of weight $\mathbf{w}=\{w_1,w_2,...,w_n\}$. Each node $i$ samples $k_i^{q}$ marbles out of group $q$, for a total of marbles equal to its own degree $K_i$. Our aim here is to show that the bias in the expected value of the covariance can be completely removed by opportunely weighing the original binary vectors. Thus, re-normalizing the vectors leads to the definition of a new covariance estimator, $\hat{\sigma}_{ij}^\mathbf{w}$, which possesses the desirable property that its expected value is zero.

Specifically, focusing on node $i$, a component $q$ of vector $\mathbf{{v}_i^w}$ is now set equal to $1/f(w_q,K_i)$ if $i$ randomly sampled a marble out of group $q$ and 0 otherwise. We can then reorder each user's weighted vector $\mathbf{{v}_i^w}$ as follows:
$$
\mathbf{{v}_i^w}=\left\{\frac{\delta_1}{f(w_1,K_i)},...,\frac{\delta_{m_1}}{f(w_1,K_i)},\frac{\delta_{m_1+1}}{f(w_2,K_i)},...,\frac{\delta_{m_1+m_2}}{f(w_2,K_i)},...,\frac{\delta_{T-m_n+1}}{f(w_n,K_i)},...,\frac{\delta_{T}}{f(w_n,K_i)} \right\},
$$
where each $\delta_s$ is either 1 or 0, and  the following constraints hold,
$$
\sum_{s=1}^{m_1}\delta_{s} ={k}_i^{1},  \cdots, \sum_{s=T-m_n+1}^{T}\delta_{s} ={k}_i^{n}; \ \ \sum_{s=1}^{T}\delta_s= \sum_{q=1}^{n}{k}_i^{q}  =K_i;  \ \ \sum_{q=1}^{n}m_q =T. 
$$

Having thus re-normilized the original vectors by the weight functions $f(w_q,K_i)$, we can now define the weighted covariance estimator as:
\begin{equation}
\hat{\sigma}_{ij}^\mathbf{w}=\frac{1}{T} \sum_{q=1}^{n} \frac{\hat{n}_{ij}^{q}}{f(w_q,K_i) f(w_q, K_j)}-\frac{1}{T^2}  \left(\sum_{q=1}^{n} \frac{{k}_i^{q}}{f(w_q,K_i)}\right) \left(\sum_{q=1}^{n} \frac{{k}_j^{q}}{f(w_q, K_j)} \right),
\label{eq:cov_w}
\end{equation}
where $\hat{n}_{ij}^{q}$ is the number of marbles of weight $w_q$ in common between $i$ and $j$.

Working under the multivariate version of the biased urn model introduced in Section \ref{sec:II}, we're now in the position to calculate the expected value of the weighted covariance. Under the Hypergeometric distribution hypothesis, see Eq. (\ref{eq:exp_nij}) we have that,
\begin{equation*}
\mathbf{E}[n_{ij}^q|k_i^1,...k_i^n,k_j^1,...k_j^n]= \frac{k_i^{q} \, k_j^{q}}{m_q },
\end{equation*}
so that the expected value of the weighted covariance in Eq. (\ref{eq:cov_w}) can be written as:
\begin{equation}
\mathbf{E}[\sigma_{ij}^\mathbf{w}]=\frac{1}{T} \sum_{q=1}^{n} \left[\frac{\mathbf{E}[k_i^{q}]}{f(w_q,K_i)} \left(\frac{\mathbf{E}[k_j^{q}]}{m_q \, f(w_q, K_j)}-\frac{1}{T}  \sum_{p=1}^{n} \frac{\mathbf{E}[k_j^{p}]}{f(w_p, K_j)} \right) \right].
\label{eq:cov_multi}
\end{equation}

From Eq. (\ref{eq:cov_multi}), we can define the group of weight functions $\{f(w_1, K_j),...,f(w_n, K_j) \}$ as those which zero the expected value of the weighted covariance, that is, the solutions of the following system of equations:
\begin{eqnarray}
\frac{\mathbf{E}[k_j^{1}]}{m_1 \, f(w_1, K_j)}&-&\frac{1}{T} \sum_{p=1}^{n} \frac{\mathbf{E}[k_j^{p}]}{f(w_p, K_j)}=0 \nonumber \\ 
\frac{\mathbf{E}[k_j^{2}]}{m_2 \, f(w_2, K_j)}&-&\frac{1}{T} \sum_{p=1}^{n} \frac{\mathbf{E}[k_j^{p}]}{f(w_p,K_j)}=0 \nonumber \\
 &\vdots& \nonumber \\
\frac{\mathbf{E}[k_j^{n}]}{m_n \, f(w_n, K_j)}&-&\frac{1}{T} \sum_{p=1}^{n} \frac{\mathbf{E}[k_j^{p}]}{f(w_p, K_j)}=0.
\label{eq:system}
\end{eqnarray}

System \ref{eq:system} is indeterminate and can be solved after assigning an arbitrary value to one of the weight functions, for example $f(w_1,K_j)$. Then all the other weight functions can be written relative to $f(w_1,K_j)$:
\begin{equation}
\frac{f(w_q,K_j)}{f(w_1, K_j)}= \frac{m_1}{m_q} \ \frac{\mathbf{E}[k_j^{q}]} {\mathbf{E}[k_j^{1}]} , \ \ \ \ \text{with} \ q \in [2,n].
\label{eq:weightfun}
\end{equation}

Thus, by defining the weight functions $\{f(w_1, k_j),...,f(w_n, k_j) \}$ with Eq. (\ref{eq:weightfun}), it's guaranteed that the expected value of the weighted covariance estimator in Eq. (\ref{eq:cov_w}) is zero.

In the multivariate case, the Wallenius distribution PDF for the vector of variables $\mathbf{k_j}=\{k_j^{1},k_j^{2},...,k_j^{n}\}$, with weight vector $\mathbf{w}=\{w_1,w_2,...,w_n\}$ and number of marbles per weight group $\mathbf{m}=\{m_1,m_2,...,m_n\}$, takes the form:
\begin{equation*}
W(\mathbf{k_j};\mathbf{m},\mathbf{w}) = \prod_{q=1}^{n} \binom{m_q}{k_j^{q}}  \int_{0}^{1} \prod_{q=1}^{n} (1-t^{w_q/D})^{k_j^{q}} \ dt, \ \ \text{with} \ \ D=\mathbf{w} \cdot (\mathbf{m}-\mathbf{k_j})=\sum_{q=1}^{n}w_q (m_q-k_j^{q}).
\end{equation*}

The group means $\mu_q=\mathbf{E}[k_j^{q}]$ with $q \in [1,n]$ satisfy the system of equations \cite{Chesson_1976}:
\begin{equation}
\left( 1- \frac{\mu_1}{m_1}\right)^{1/w_1}=\left( 1- \frac{\mu_2}{m_2}\right)^{1/w_2}= ...=\left( 1- \frac{\mu_n}{m_n}\right)^{1/w_n},
\label{eq:mu}
\end{equation}
with the constraint $\sum_{q=1}^{n} \mu_q=K_j$. From this constraint and Eq. (\ref{eq:weightfun}), we can write each group mean $\mu_q$ in terms of the weight functions,
\begin{equation}
\frac{\mu_q}{m_q}=\frac{ K_j \, f(w_q,K_j)}{\sum_{p=1}^{n} m_p \, f(w_p,K_j)},
\label{eq:mu_f}
\end{equation}
and inserting Eq. (\ref{eq:mu_f}) in Eq. (\ref{eq:mu}), we find a set of equations for the weight functions:
\begin{equation}
\left(1-\frac{k_j \, f(w_1,k_j)}{\sum_{p=1}^n m_p \, f(w_p,k_j)} \right)^{1/w_1}= ...=\left(1-\frac{k_j \, f(w_n,k_j)}{\sum_{p=1}^n m_p \, f(w_p,k_j)} \right)^{1/w_n}.
\label{eq:weight_f_eq}
\end{equation}

System \ref{eq:weight_f_eq} provides a way to directly calculate the weight functions, without having to compute the group means first.

\section{\label{sec:IV}Multivariate weighted correlation estimator}

In this section, we write down the weighted estimator for the correlation coefficient and quantitatively show the improvement it offers over the unweighted one.

From Eq. (\ref{eq:cov_multi}) it's straightforward to define the weighted correlation coefficient estimator as the Pearson correlation coefficient of the weighted vectors:
\begin{eqnarray}
& & \hat{\rho}_{ij}^\mathbf{w}=\frac{\hat{\sigma}_{ij}^\mathbf{w}}{\hat{\sigma}_{i}^\mathbf{w} \ \hat{\sigma}_{j}^\mathbf{w}}= \nonumber \\
&=& \frac{ \sum_{q=1}^{n} \frac{{n}_{ij}^q}{f(w_q,K_i) f(w_q, K_j)} -\frac{1}{T} \left( \sum_{q=1}^{n} \frac{{k}_i^{q}}{f(w_q, K_i)} \right) \left(\sum_{q=1}^{n} \frac{{k}_j^{q}}{f(w_q, K_j)} \right) }{\sqrt{\left[\sum_{q=1}^{n} \frac{{k}_i^{q}}{f(w_q,K_i)^2}-\frac{1}{T} \left(\sum_{q=1}^{n} \frac{{k}_i^{q}}{f(w_q,K_i)} \right)^2 \right]  \left[\sum_{q=1}^{n} \frac{{k}_j^{q}}{f(w_q,K_j)^2}-\frac{1}{T} \left(\sum_{q=1}^{n} \frac{{k}_j^{q}}{f(w_q, K_j)} \right)^2 \right]}}.
\label{eq:cor_weighted}
\end{eqnarray}

Unfortunately, from Eq. (\ref{eq:cor_weighted}) one realizes immediately that having $\mathbf{E}[\sigma_{ij}^\mathbf{w}]=0$ is not a sufficient condition for $\mathbf{E}[\rho_{ij}^\mathbf{w}]=0$, since variables $\{\mathbf{k_i},\mathbf{k_j} \}$ now appear in the denominator as well. However, we can approximate $\mathbf{E}[\rho_{ij}^\mathbf{w}]$ by its Taylor series near $\mathbf{w}=\mathbf{1}$ and show that its value is less than the Taylor series of $\mathbf{E}[\rho_{ij}]$.

\subsection{Comparison of correlation coefficients near $\mathbf{w}=\mathbf{1}$}
We now proceed to show the improvement of the weighted estimator over the unweighted one, by comparing the Taylor series of their expected values. Out of simplicity, we show our results in the bivariate case, with $n=2$ groups and $w=w_2/w_1$. The Taylor series of $\mathbf{E}[\rho_{ij}]$ near $w=1$ was calculated in Section \ref{sec:II}, Eq. (\ref{eq:approx_rhoij}).

We now calculate the Taylor series of $\mathbf{E}[\rho_{ij}^w]$, starting from the expected value of $\rho_{ij}^w$ given $k_i^w,k_j^w$, which can be calculated from Eq. (\ref{eq:cor_weighted}) when $n=2$:
\begin{equation}
\mathbf{E}[\rho_{ij}^w|k_i^w,k_j^w]= \frac{\left[(T-m) \, k_i^w - m \, f(w,K_i)(K_i-k_i^w) \right]}{ m \,T \, \sigma_i^w f(w,K_i) } \frac{ \left[(T-m) \, k_j^w- m \, f(w,K_j)(K_j-k_j^w) \right]}{ (T-m) \, T \, \sigma_j^w f(w,K_j)}.
\label{eq:Wall_rho_dec}
\end{equation}

From Eq. (\ref{eq:Wall_rho_dec}), remembering that the Wallenius distribution in $w=1$ becomes the Hypergeometric distribution, we can calculate the zero order term in the Taylor series, which turns out to be null. To calculate the first and second order terms, we define the function:
$$
 F(k_i^w,k_j^w,w)=\mathbf{E}[\rho_{ij}^w|k_i^w,k_j^w] \cdot W(k_i^w) \cdot W(k_j^w),
$$
which, summed over all possible values of $\{k_i^w,k_j^w \}$ gives $\mathbf{E}[\rho_{ij}^w]$. Thus, we can calculate the derivatives as follows,
\begin{equation*}
\left. \frac{d\mathbf{E}[\rho_{ij}^w]}{dw} \right|_{w=1}= \sum_{k_i^w,k_j^w} \left[  \frac{d }{d w}  \mathbf{E}[\rho_{ij}^w|k_i^w,k_j^w] W(k_i^w) W(k_j^w) \right]_{w=1}=\sum_{k_i^w,k_j^w} \left. \frac{dF(k_i^w,k_j^w,w)}{dw} \right|_{w=1},
\end{equation*}
by exploiting the advantage of first evaluating the derivatives of $F(x_i,x_j,w)$ near $w=1$, and then summing over the variables. The first non-null term is the second order one, so that the expected value of the weighted correlation coefficient near $w=1$ is:
\begin{eqnarray}
\mathbf{E}[\rho_{ij}^w]  & \simeq  &  \frac{m(T-m)}{T\sqrt{K_i  \left(1-\frac{K_i}{T} \right) K_j  \left(1-\frac{K_j}{T} \right)}}  \left( 1-\frac{K_i}{T} \right) 
 \left[h_{(T)}-h_{(T-K_i)}+\left(1-\frac{1}{K_i}\right) \ln \left(1-\frac{K_i}{T}\right)\right]  \cdot \nonumber \\
& \cdot &\left( 1-\frac{K_j}{T} \right) \left[h_{(T)}-h_{(T-K_j)}+\left(1-\frac{1}{K_j}\right)  \ln \left(1-\frac{K_j}{T}\right)\right](w-1)^2,
\label{eq:weighted_Wall}
\end{eqnarray}
where $h_{(n)}=\sum_{k=1}^n 1/k$ is the $n$-th harmonic number, that is, the sum of the reciprocals of the first $n$ natural numbers.

\begin{figure}[!htb]
    \includegraphics[width=0.9\textwidth]{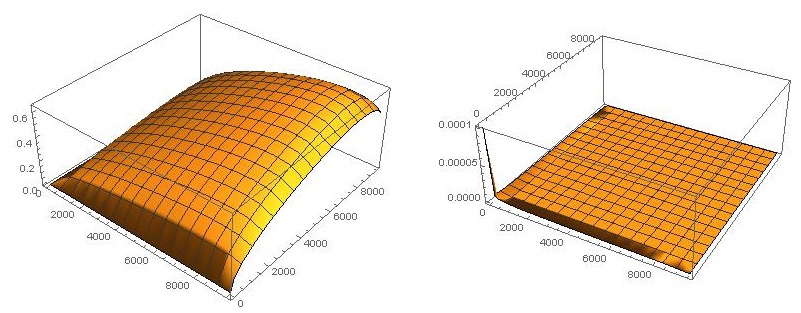}
    \caption{Plot of the expected value of the unweighted correlation coefficient (left) against the weighted one (right) as a function of $k_i$ and $k_j$. Parameters are: $T=10^4=2m$; $w=2$; $\{k_i,k_j\} \in [1, 0.95T]$. Both assume the same value of $0.0001$ in $\{1,1\}$. Notice that the vertical scales are different in the left and right plots.}
    \label{fig:cor}
\end{figure}

A graphic comparison between the unweighted estimator in Eq. (\ref{eq:approx_rhoij}) and the weighted estimator in Eq. (\ref{eq:weighted_Wall}) is shown in Fig \ref{fig:cor}, where the improvement of the latter is clear.

Finally, to quantify the improvement offered by the weighted estimator over the unweighted one, we use the asymptotic expansion of the harmonic number,
\begin{equation*}
h_{(T)}-h_{(T-K_i)} \simeq  -\ln \left(1-\frac{K_i}{T}\right)-\frac{1}{2T} \left(\frac{K_i/T}{1-K_i/T} \right),
\end{equation*}
valid when $T \rightarrow \infty$ and $T>>K_i$.

Within the former asymptotic limit, we have that the ratio of the expected value of the weighted correlation coefficient to the unweighted one, near $w=1$, is
\begin{eqnarray}
\frac{\mathbf{E}[\rho_{ij}^w]}{\mathbf{E}[\rho_{ij}]}  &=& 
\left[\frac{h_{(T)}-h_{(T-K_i)}}{\ln \left(1-\frac{K_i}{T}\right)}+1-\frac{1}{K_i}\right] \left[\frac{h_{(T)}-h_{(T-K_j)}}{\ln \left(1-\frac{K_j}{T}\right)}+1-\frac{1}{K_j}\right] \simeq \nonumber \\
& \simeq & \left(\frac{1}{K_i} -\frac{1}{2T}\right) \left(\frac{1}{K_j} -\frac{1}{2T} \right)\simeq \frac{1}{K_i K_j}.
\end{eqnarray}

Thus, when $T>>K_i,K_j$, we find that the expected value of the weighted correlation estimator is $1/K_i K_j$ times the expected value of the unweighted one. 

\section{\label{sec:V} Empirical datasets}

In this section, we employ the weighted covariance and correlation estimators we developed, against the unweighted ones, with the aim of showing how the new estimators get rid of the noise present in the rewired network. As a matter of fact, in order to calculate the weighted covariance and correlation, we simply derive the weight functions as shown in section \ref{sec:III} and use them to weigh users' vectors, over which we then compute the covariance and correlation coefficient.

The datasets taken into consideration are two, one pertains to the social sciences and the other one to the biological sciences. The social database \cite{Puccio_PhysA_2016} consists of 1,808 private initiatives submitted between 2011 and 2014 by 199 members of the Finnish parliament (MPs), along with information on who signed each initiative. Data cover an entire parliament of the duration of four years. The resulting bipartite system displays MPs on one side and initiatives they signed on the other. Info on MPs include their party and district of election. Parties in Finland are: Christian Democrats (KD), Centre party (KESK), National Coalition party (KOK), Finns party (PS), Swedish People's party (RKP), Social Democratic party (SDP), Left alliance (VAS) and Green League (Vihr). Electoral districts are 15.

The biological data comes from the COG database (\footnote{available at http://www.ncbi.nlm.nih.gov/COG}), which stands for Clusters of Orthologous Groups of proteins, from the sequenced genomes of prokaryotes and unicellular eukaryotes. The database consists of 4,873 COGs present in 66 genomes of unicellular organisms, belonging to 3 broad macro-groups: Archaea, Bacteria or Eukaryota. The corresponding bipartite system consists of organisms on one side and COGs present in their genome on the other. Organisms belong to 12 different phyla: Actinobacteria (Act), Archaea of type Crenarchaeota (ArC) and Euryarchaeota (ArE), Cyanobacteria (Cya), Eukariota (Euk), Gram-negative Proteobacteria of type $\alpha$ (Gr-a), $\beta$ (Gr-b), $\epsilon$ (Gr-e), $\gamma$ (Gr-g), Gram-positive bacteria (Gr+), Hyperthermophilic bacteria (HyT) and other bacteria (Oth). This database has been widely studied, see for example  \cite{COGS1} and \cite{COGS2}.

The fundamental property in both datasets that makes them well-suited for our purpose is the high degree of heterogeneity present in both sides of the bipartite system, as can be seen from TABLE \ref{tab:data}. However, such a high degree of heterogeneity is frequently found in bipartite systems.
\begin{table}
\begin{ruledtabular}
\begin{tabular}{c | l  l  l  l } 
\multicolumn{3}{c}{\textbf {Data}}  \\ \toprule
 & \textit{Finnish parliament} & \textit {COGS} \\
\hline
$T$ & 1,808 & 4,873 \\
$w_m-w_M$ & 2-150 & 3-66 \\
$N$ & 199 & 66\\
$K_m-K_M$ & 2-793 & 362-2,243\\
$n_L$ & 32,024 & 83,675 \\
\end{tabular}
\end{ruledtabular}
\caption {\label{tab:data} $T$ is the number of initiatives/COGs; $w_m-w_M$ is their heterogeneity, that is, the range (min-max) of degree distributions; $N$ is the number of MPs/organisms; $K_m-K_M$ is the range (min-max) of their degree distributions; $n_L$ is the number of links in the bipartite network.}
\end{table}

\subsection{Rewiring algorithm}
If we want to assess the noise present in the correlation matrix computed according to Eq. (\ref{eq:Pearson_c}), one of the approaches used in the literature \cite{Vespignani_NatPhys_2006} is the rewiring of the bipartite network. Our rewiring algorithm samples randomly a pair of MPs/organisms according to a probability distribution equal to their degree distribution, then samples randomly two initiatives/COGs out of those already linked to the first sampled pair, again according to the degree distribution of initiatives/COGs. Then, if neither in the pair is already linked to the other's sampled initiative/COG, the two links are swapped, otherwise the swap is rejected. The swapping procedure is iterated many times, until full randomization is achieved. Such an algorithm performs a random rewiring of the entire bipartite system, preserving both sides degree distributions. 

\begin{figure}[!htb]
	\includegraphics[width=0.49\textwidth]{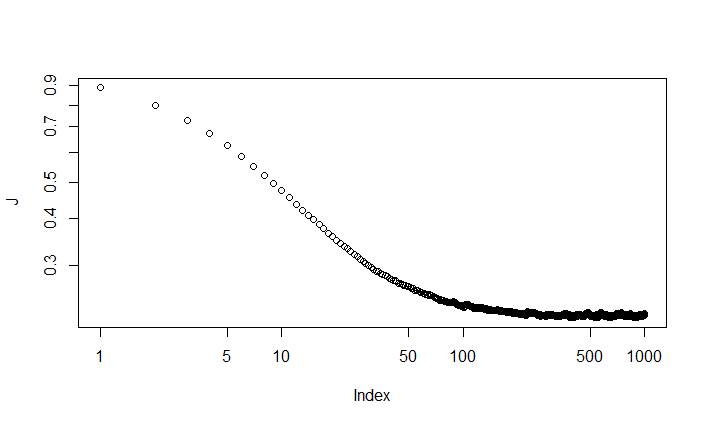}
	\includegraphics[width=0.49\textwidth]{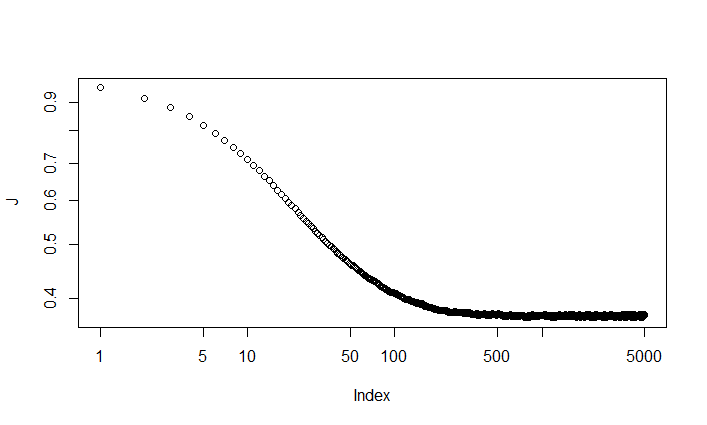}
    \caption{Jaccard index as a function of the number of shuffles for the rewired social (left) and biological (right) networks, in log-log scale. A unit in the x-axis corresponds to 1,000 shuffles.}
    \label{fig:Jaccard}
\end{figure}

\begin{figure}[!htb]
	\includegraphics[width=0.49\textwidth]{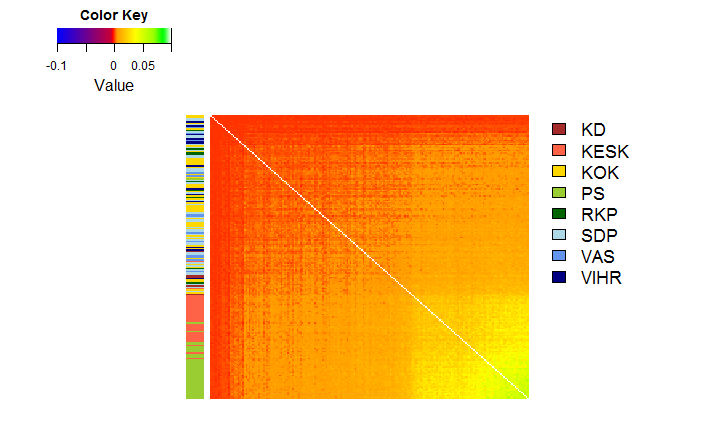}
	\includegraphics[width=0.49\textwidth]{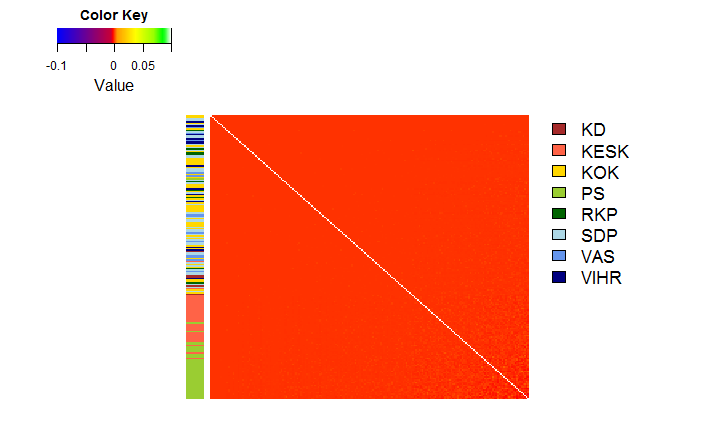}
	\includegraphics[width=0.49\textwidth]{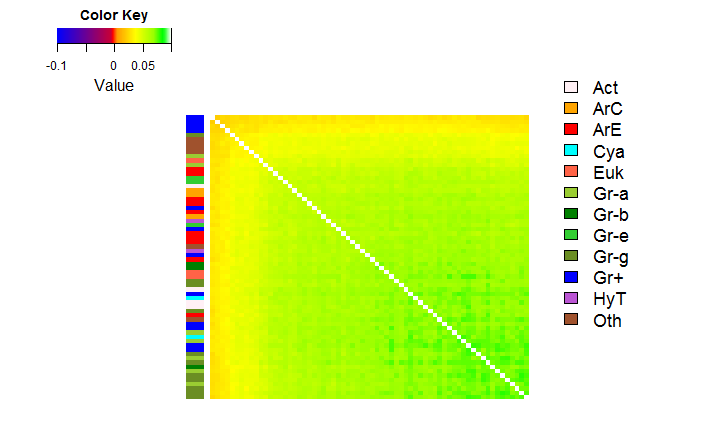}
	\includegraphics[width=0.49\textwidth]{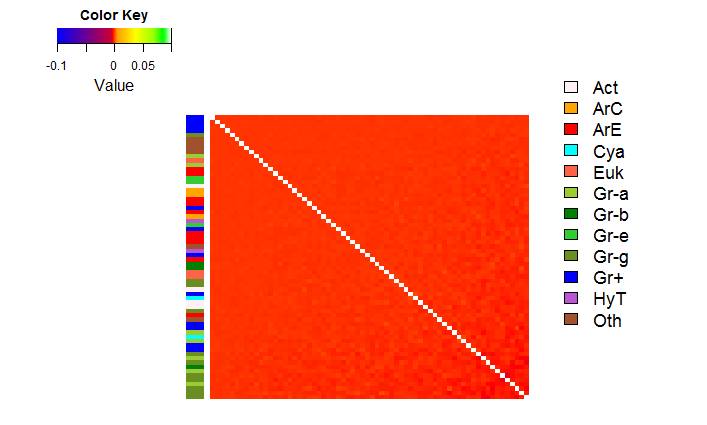}
    \caption{Covariance matrices of MPs (top-row) and organisms (bottom-row) after random rewiring of the original bipartite network, calculated without weighing the vectors (left) and weighing them (right). MPs/organisms are ordered by increasing degree to better show the bias and diagonals have been colored white. The Color Key scale is identical in all figures.}
    \label{fig:rew_cov}
\end{figure}

\begin{figure}[!htb]
	\includegraphics[width=0.49\textwidth]{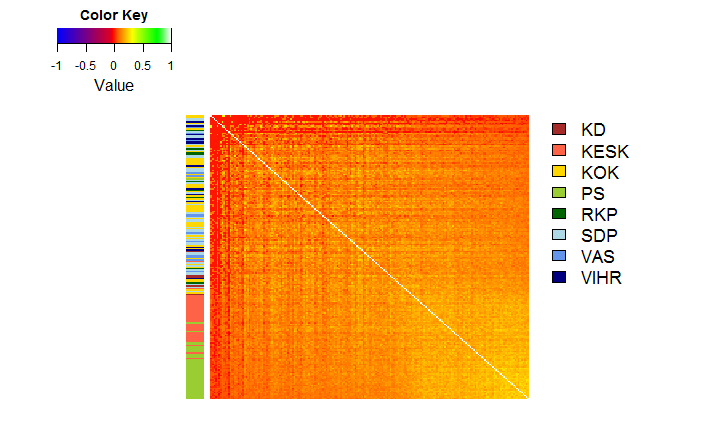}
	\includegraphics[width=0.49\textwidth]{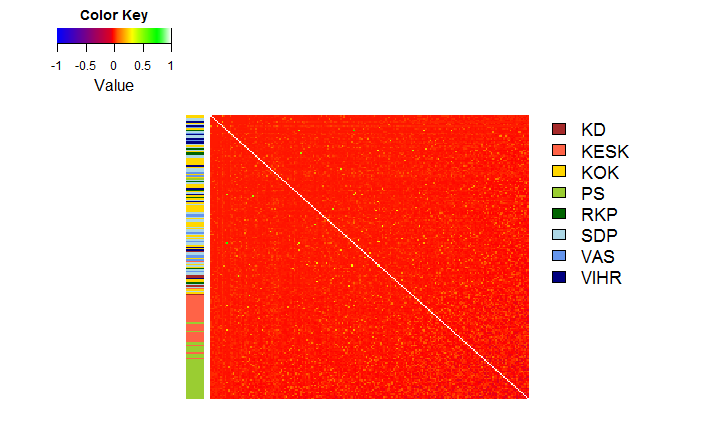}
	\includegraphics[width=0.49\textwidth]{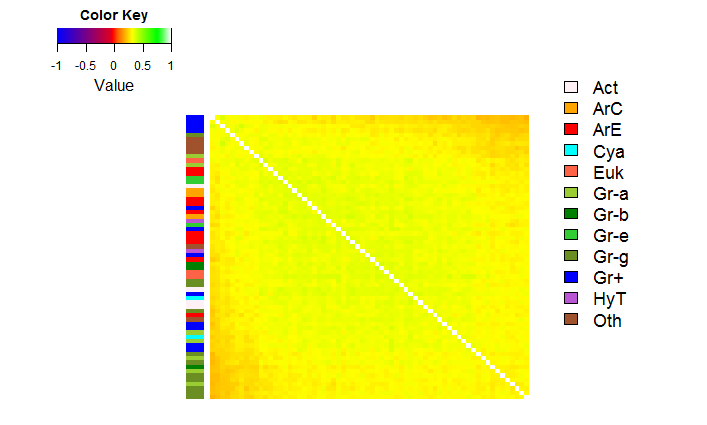}
	\includegraphics[width=0.49\textwidth]{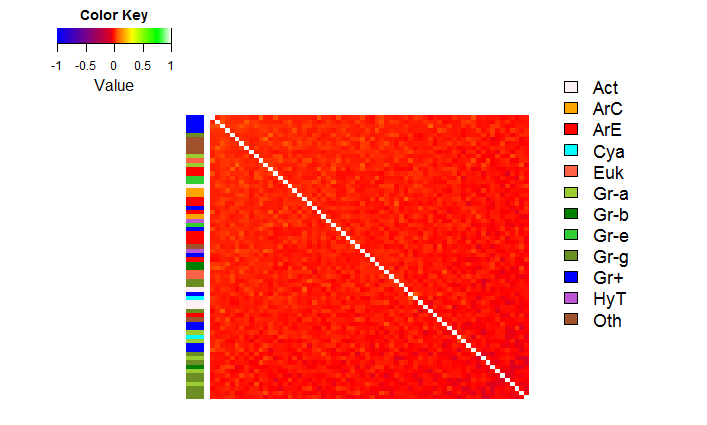}
    \caption{Correlation matrices of MPs (top-row) and organisms (bottom-row) after random rewiring of the original bipartite network, calculated without weighing the vectors (left) and weighing them (right). MPs/organisms are ordered by increasing degree to better show the bias and diagonals have been colored white. The Color Key scale is identical in all figures.}
    \label{fig:rew}
\end{figure}

\begin{figure}[!htb]
	\includegraphics[width=0.49\textwidth]{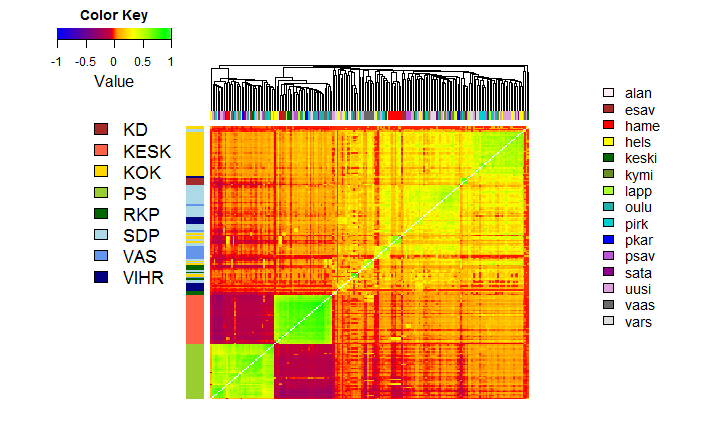}
	\includegraphics[width=0.49\textwidth]{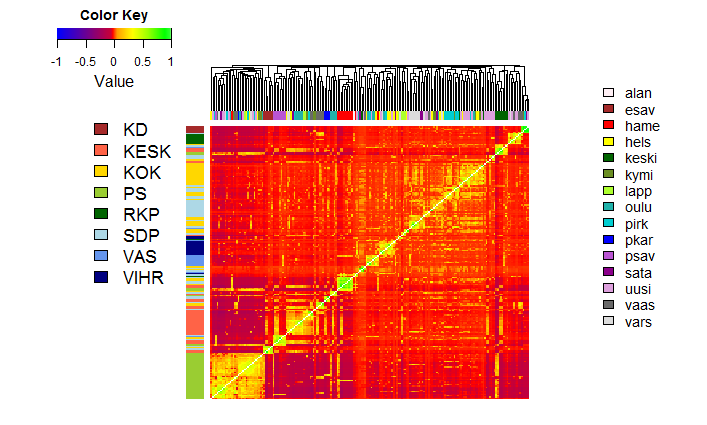}
	\includegraphics[width=0.49\textwidth]{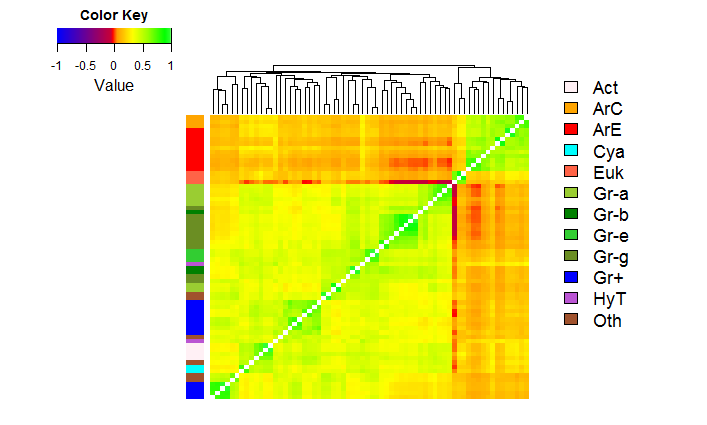}
	\includegraphics[width=0.49\textwidth]{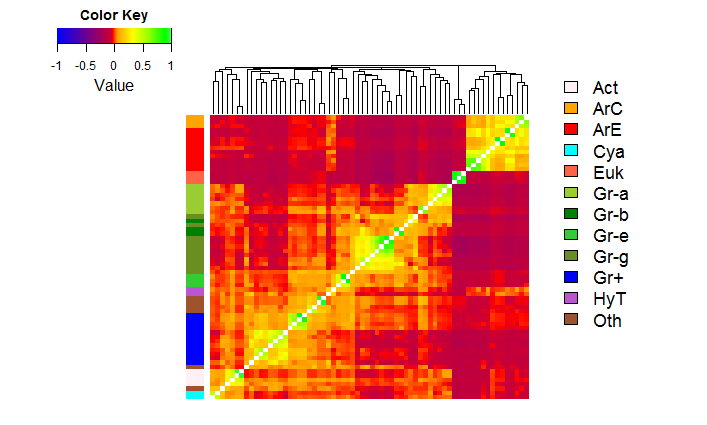}
    \caption{Unweighted (left) against weighted (right) correlation matrices of MPs (top) and organisms (bottom), ordered by hierarchical clustering with average linkage performed on each matrix \cite{Anderberg}. The left-side bar is colored according to party (left legend) or phylum (right legend), the top bar is colored according to districts (right legend). Diagonals have been colored white. The Color Key scale is identical in all figures.}
    \label{fig:differences}
\end{figure}

In the former procedure, the only problem lies in understanding when to stop, that is, we need a measure of randomness. We employed the Jaccard similarity index \cite{Jaccard_1912}  as a measure of randomness, terminating the algorithm after one million shuffles for the social dataset and after five million for the biological dataset, well after the Jaccard index stabilised around a minimum (see Fig. \ref{fig:Jaccard}).
We can now compare the weighted estimators against the unweighted ones, over both datasets. The first result, as shown in Fig. \ref{fig:rew_cov}, is that the weighted covariance estimator completely destroys the structure still present in the unweighted covariance matrix of the rewired network. This feature translates also to the weighted/unweighted correlation coefficients in Fig. \ref{fig:rew}, although, where the expected value of the weighted covariance estimator is zero, we only have an approximated result for the expected value of the weighted correlation estimator.
In Fig. \ref{fig:differences} we show how the weighing, though getting rid of the noise, still grasps the cluster-structure present in the system. 

An unexpected result is that the weighted correlation matrix seems to actually better identify the clusters in the COGs dataset (bottom row), by encompassing a broader scale of values, displayed within the matrix in violet (negative correlations), zero (red), orange (low), yellow (average) and green (high) against the unweighted matrix which only features the positive correlations, making it harder to distinguish subclusters. Indeed the right weighted matrix shows subclustering corresponding to organisms' phyla. For example, it neatly discriminates Archaea (red and orange in the left color-bar), Eukariota (Salmon) and Bacteria (all the rest), by also grouping together Gram-negative bacteria (shades of green), Gram-positive bacteria (blue), Hyperthermophilic bacteria (violet), Actinobacteria (pink) and Cyanobacteria (cyan).

In the Finnish parliament dataset, in Fig. \ref{fig:differences} top row, we can see how the weighing destroys the cluster of party KESK, showing how this cluster is more due to noise than to a real collaboration between MPs, while at the same time it preserves the cluster of party PS, which appears to be a real one. This finding is in accord with the general trend observed in \cite{Puccio_PhysA_2016}, where the evolution of this network over 4 Finnish parliament terms is studied. In fact, during previous terms, MPs collaborated by district and by party both, with party beeing more charcterizing in the opposition and district subclustering within the government. If we look at the unweighted matrix, it appears that not only the two opposition parties strongly cluster and display a negative correlation with each other, but also the government splits in two right-wing left-wing subclusters. Such a change from the previous terms was attributed to the sudden rise in numbers of the populist party PS. From the weighted matrix instea we can see that the situation is more in line with previous terms, with district subclustering reappearing.

\subsection{Robustness analysis}
Since the proposed weighted estimator depends on both sides degrees, if we sample a subset of the group of interest (MPs/organisms), the degree distribution on the other side (initiatives/COGs) decreases as well and, as a result, the weighted correlations change for the remaining elements in the set of interest. Since this makes it so that the correlation coefficient between two elements would depend on who is also present in the subset, a robustness analysis is in order, to show how the weighted estimator holds up when subsetting data. 

We ran 1,000 sampling of, respectively, 90\%, 80\% and 70\% MPs/organisms out of the rewired network, then plotted the Frobenius distance between pairs of weighted correlation matrices (by taking each time the intersection of those present in both matrices), the Frobenius distance between the 1,000 sampled weighted correlation matrices and the identity matrix (which corresponds to the noiseless null-model) and the Frobenius distance \cite{Horn} between the 1,000 sampled unweighted correlation matrices and the identity matrix. In order to compare matrices of different dimensions, we renormalized each distance by $\sqrt{n(n-1)}$, where $n$ is the size of the pair of matrices over which the distance is calculated.
\begin{figure}[!htb]
	\includegraphics[width=0.32\textwidth]{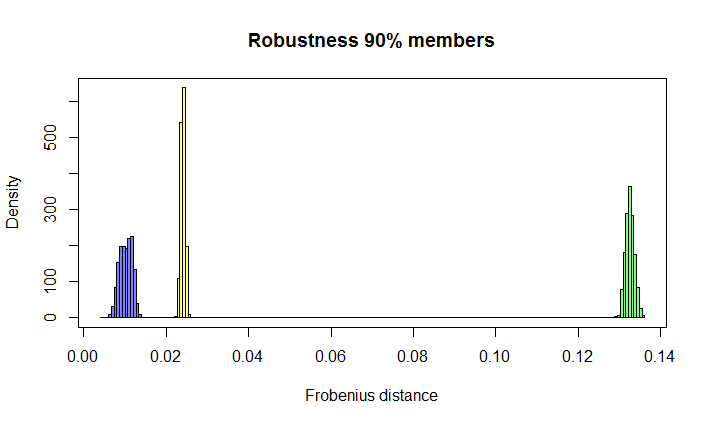}
	\includegraphics[width=0.32\textwidth]{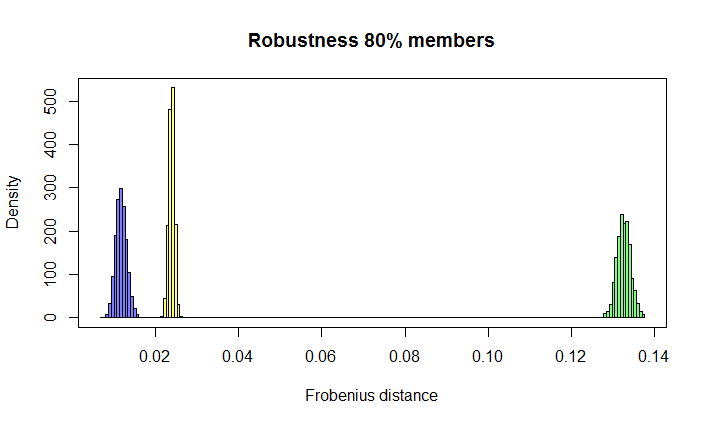}
	\includegraphics[width=0.32\textwidth]{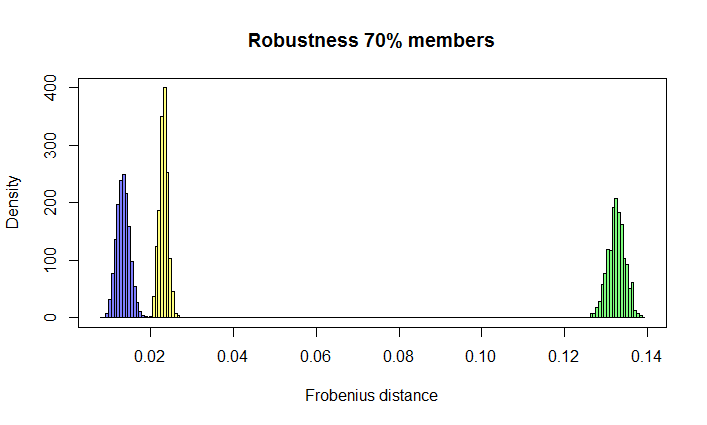}
	\includegraphics[width=0.32\textwidth]{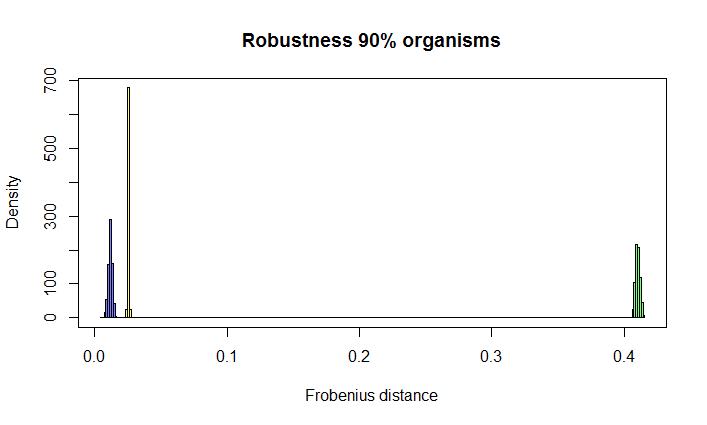}
	\includegraphics[width=0.32\textwidth]{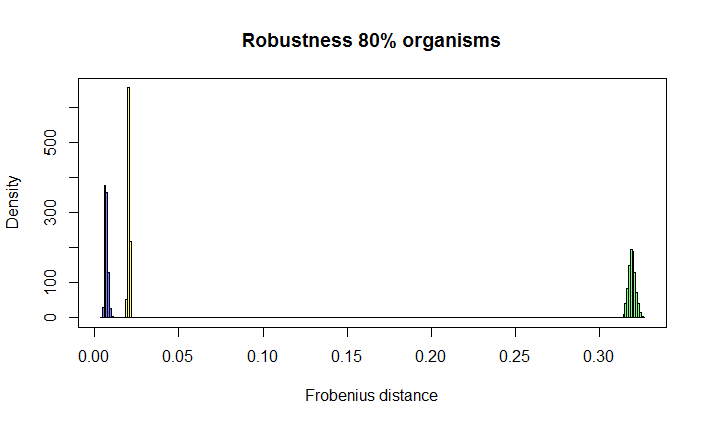}
	\includegraphics[width=0.32\textwidth]{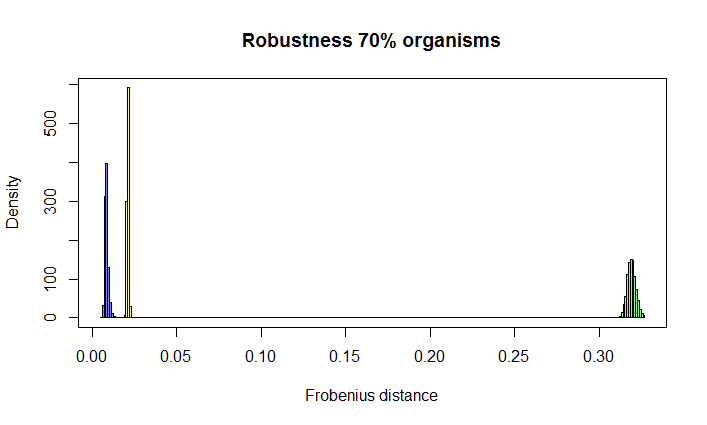}
    \caption{Robustness analysis performed on the weighted correlation coefficient between MPs (top) and between organisms (bottom) in the rewired network. We display in violet the distribution of Frobenius distances between weighted correlation matrices, in yellow the distribution of weighted-Identity distances, in green the distribution of unweighted-Identity distances.}
    \label{fig:rob}
\end{figure}
From Fig. \ref{fig:rob} we can clearly see that, although the distribution of distances between weighted correlation matrices broadens and shifts to the right as the percentage of sampled elements decreases, it still stands to the left of the weighted-Identity distance distribution, which keeps well to the left of the unweighted-Identity one. We can thus conclude that the weighted estimator proves robust even when sampling subsets of data.

\subsection{\label{odds}Weight-groups and Odds-ratios Estimation}

Up to now we've run all the calculations under the following assumption: the odds-ratios estimator $\mathbf{w}$ is exactly equal to the heterogeneity of set B in the bipartite system, meaning, for example, that in the Finnish parliament the weight of an initiative is equal to the number of MPs who signed it and in the COGs dataset, the weight of a COG is equal to the number of organisms in whose genome it shows up. Such a rough estimate has the benefit of automatically defining the weight-groups vector $\mathbf{m}$ as well, by grouping together all the initiatives/COGs which have the same weight.

In truth, the estimation of the odds-ratios in a Wallenius distribution with different sampling processes, that is, a different number of total marbles sampled by each user, is not straightforward in and of itsels, and has not been thoroughly investigated in the literature. In this section, we propose a simple method, whose efficacy is proven through simulations, of determining first the groups of marbles of different weight and then the weight corresponding to each group. In this context, we'll show the improvement the weighted estimators for both covariance and correlation offer over the simple idea of just dividing the original vectors $\mathbf{v_i}$ $(\mathbf{v_j})$ by the weight $\mathbf{w}$ defined by set B's heterogeneity, as inspired by the reading of Newman's paper \cite{Newman_2}, which shall henceforth be referred to as Newman's estimator. Basically, Newman's estimator may work well when one is dealing with datasets with low hetereogeneity, so that the noise can be modeled as a multinomial distribution, but it becomes dramatically biased as heterogeneity on both sides of the system grows, as is typically the case in many complex systems.

The setting of the simulation is as follows: we define set A heterogeneity, by fixing $\mathbf{v_i}$'s degree for every $i$, we consider four groups of marbles of equal size, and set the odds-ratios as $\mathbf{w}=\{0.05, 0.2, 0.5, 1\}$, since all the weights can be normalized in terms of any of the other weights. We ran an exploratory simulation with $\mathbf{m}=\{1000,1000,1000,1000\}$, encompassing the whole spectrum of values of $K_i$, from 10 to 3990 in steps of 10 for a total of 300 users. With these initial parameters, the simulation runs a random sampling from a biased urn with the odds-ratios $\mathbf{w}$, one user at a time, then the marbles sampled in each weight-group are labeled randomly out of the total set of labeled marbles, so that the corresponding user's profile binary vector can be constructed. Finally, the adjacency matrix is built by just stacking all profile vectors one next to the other, taking care of removing any marbles in the biased urn which were never sampled by any user.

Having thus constructed our synthetic database, we can easily calculate Newman's covariance and correlation estimators by simply dividing every row of the matrix by its corresponding marble's weight, which is just the number of users who sampled it, and then computing the unweighted estimators on the resulting matrix. 

For what concerns our newly proposed weighted estimators, in order to calculate the weight functions $f(w_h,K_i)$ one needs to estimate both the weight-groups $\mathbf{m}$ and the odds-ratios $\mathbf{w}$ from the synthetic dataset. In order to do so, we suggest a 3 steps procedure:

\begin{enumerate}
\item Calculate the distribution of the marbles in weight-groups by

\begin{enumerate}
\item counting the number of users who sampled each marble, that is the sum of each row of the binary matrix (the row-sum),
\item tallying all the marbles (the rows in the matrix) sampled by the same number of users (all those with the same row-sum), which is simply the distribution of the marbles in weight-groups, and plotting this distribution as a $xy$ curve, where the $y$ is the tally (number of marbles with a given row-sum) and the $x$ is the corresponding row-sum.
\end{enumerate}

\item The main issue is how to identify the groups under the obtained $xy$ curve, a task accomplished, for example, by using a segmented algorithm on a linear fit of the curve, to determine the points $psi$ where the slope of the curve changes. These breaking points identify the group of marbles in a given row-sum interval, so that one can calculate the vector $\mathbf{m}=\{m_1,m_2,\dots,m_n\}$ and which marbles belong to each group. Notice that each group needs to contain more than one point of the $xy$ curve, in order for the odds-ratios equation detailed at the next step to work, and it's up to one whether to use left closed or right closed intervals.

\item Once one knows the groups of marbles, the odds ratio can be estimated by setting the heaviest group weight equal to one: $w_n=1$, and then building a vector of $n-1$ weights for each user, according to the following equation inspired by Eq. (\ref{eq:mu}):

\begin{equation}
w^i_{q}=\frac{\ln \left(1-k^i_q/m_q \right)}{\ln \left(1-k^i_n/m_n \right)}.
\label{eq:oddsratio}
\end{equation}

From Eq. (\ref{eq:oddsratio}) it's possible to reconstruct each weight by averaging over all the users and keeping in mind that, in a multivariate Wallenius distribution, the odds-ratios are distributed according to a log-normal:

\begin{equation}
\left< w_{q} \right>=\exp \left( \left< \ln \left(w_q^i \right) \right>_i \right).
\label{eq:oddsratio2}
\end{equation}

The odds-ratios estimates obtained from Eq. (\ref{eq:oddsratio2}) get more and more accurate as the number of users and marbles in each group grows and the better one identifies the weight-groups. Obviously, when going from Eq. (\ref{eq:oddsratio}) to Eq. (\ref{eq:oddsratio2}), one needs first to remove all the values of $w^i_{q}$ that are either $0$, $1$ or infinite.
\end{enumerate}

\begin{figure}
\center
	\includegraphics[width=0.49\textwidth]{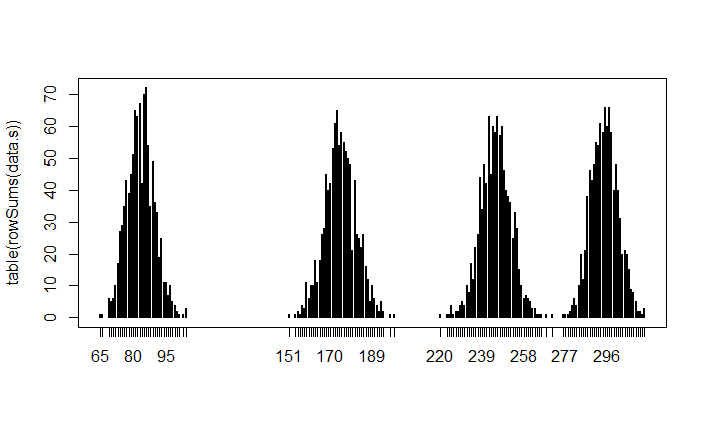}
	\includegraphics[width=0.49\textwidth]{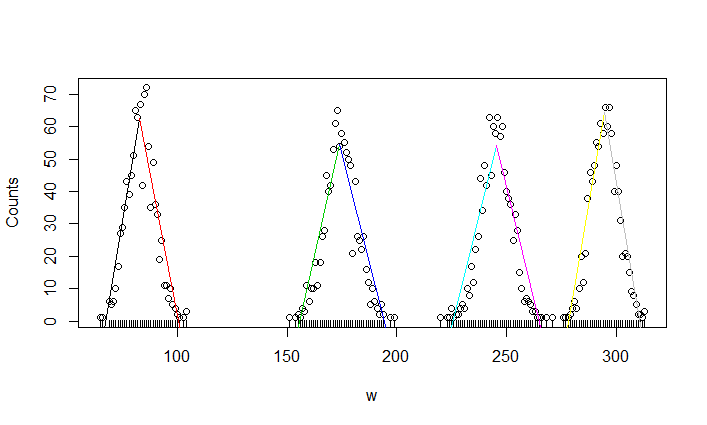}
	\includegraphics[width=0.49\textwidth]{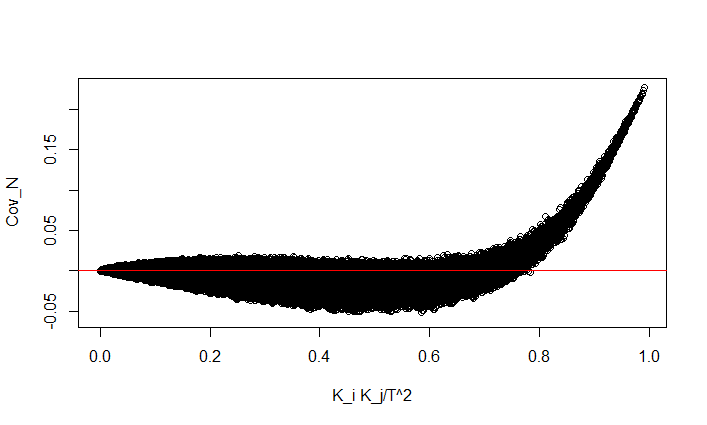}
	\includegraphics[width=0.49\textwidth]{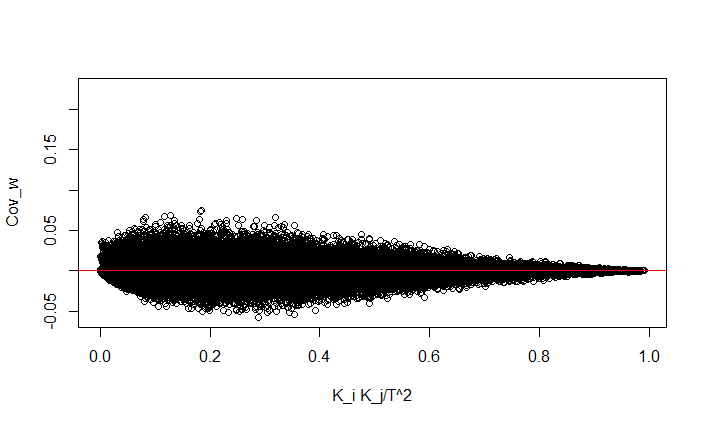}
	\includegraphics[width=0.49\textwidth]{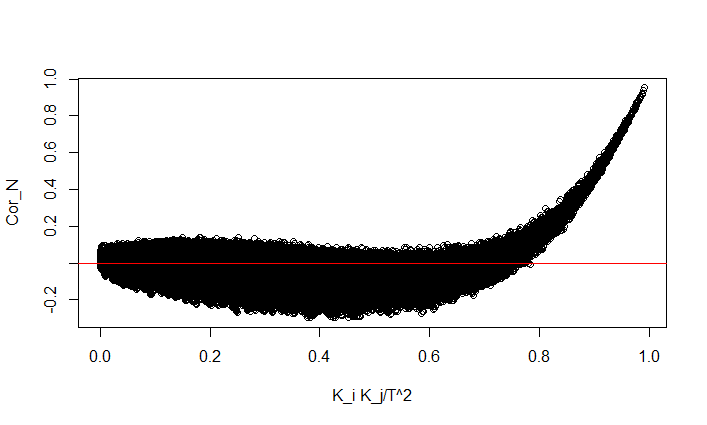}
	\includegraphics[width=0.49\textwidth]{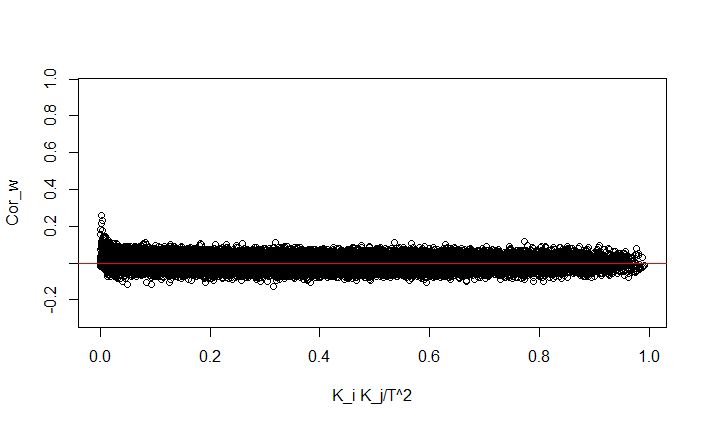}
    \caption[Weight-groups and odds-ratios of exploratory synthetic data]{Exploratory simulation, top row shows the ditribution of the row-sums of the data matrix and the segmented fit used to identify the weight-groups, mid row shows the plot of Newman's covariance (left) and weighted covariance (right) as a function of $K_i K_j/T^2$ and the bottom row shows the same plot of Newman's correlation and weighted one.}
    \label{fig:expl}
\end{figure}

In Fig. \ref{fig:expl} we report the results of the exploratory simulation, by showing the distribution of the row-sums of the synthetic data matrix, the identification of weight-groups through a linear segmented algorithm, the plot of both covariance and correlation estimators calculated with Newman's weight and with our weighted estimators as a function of users' degree: $K_i K_j/T^2$, $\forall i,j>i$. 

\begin{figure}
\center
	\includegraphics[width=0.49\textwidth]{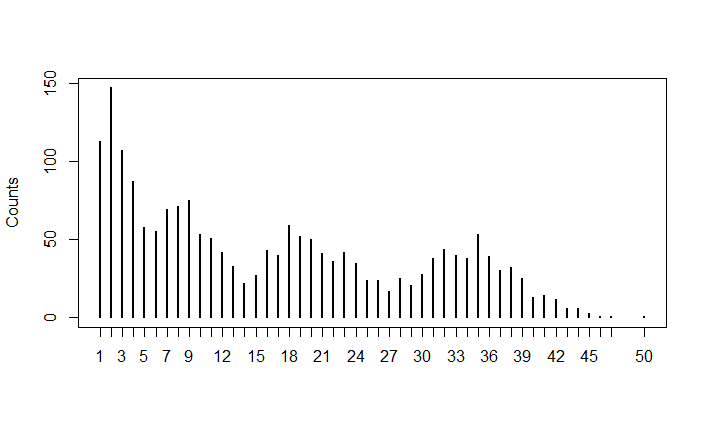}
	\includegraphics[width=0.49\textwidth]{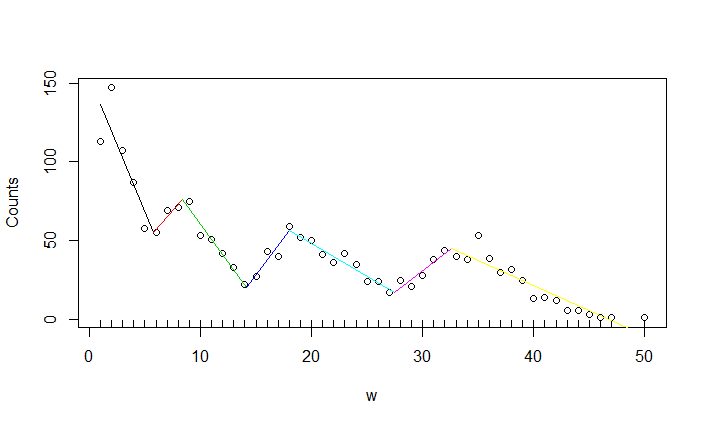}
	\includegraphics[width=0.49\textwidth]{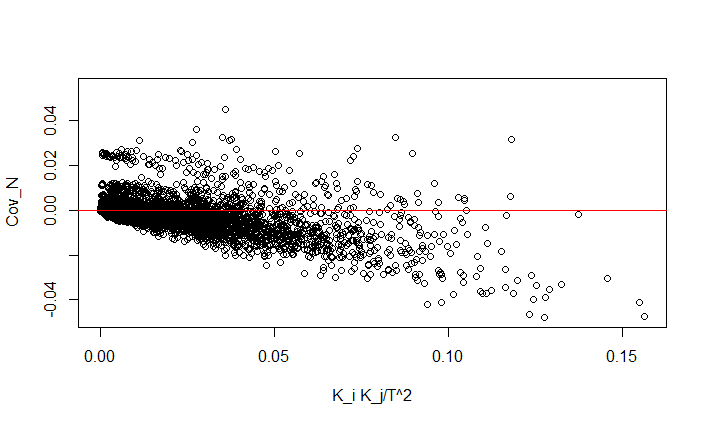}
	\includegraphics[width=0.49\textwidth]{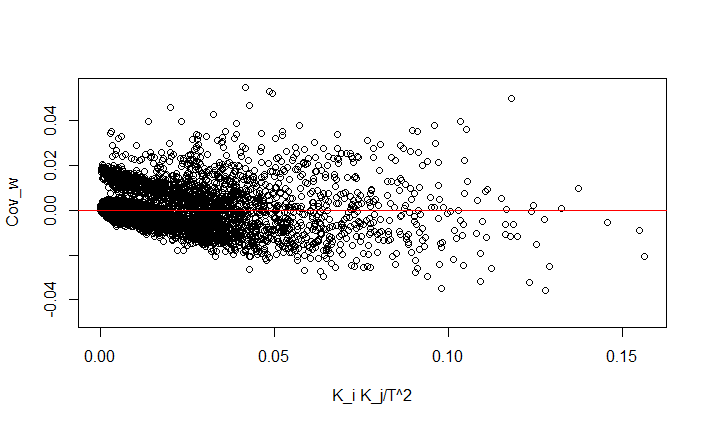}
	\includegraphics[width=0.49\textwidth]{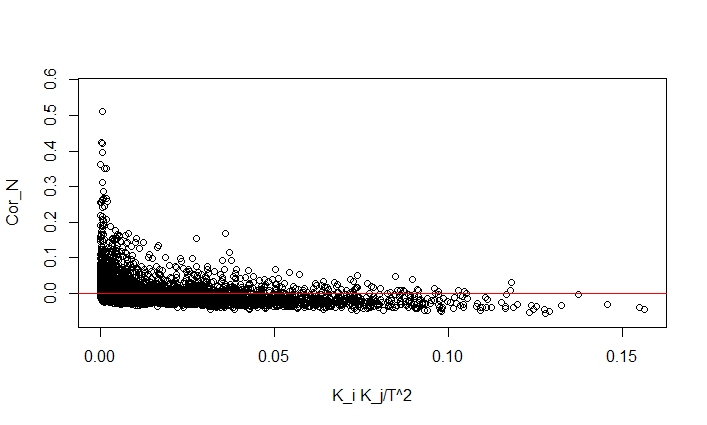}
	\includegraphics[width=0.49\textwidth]{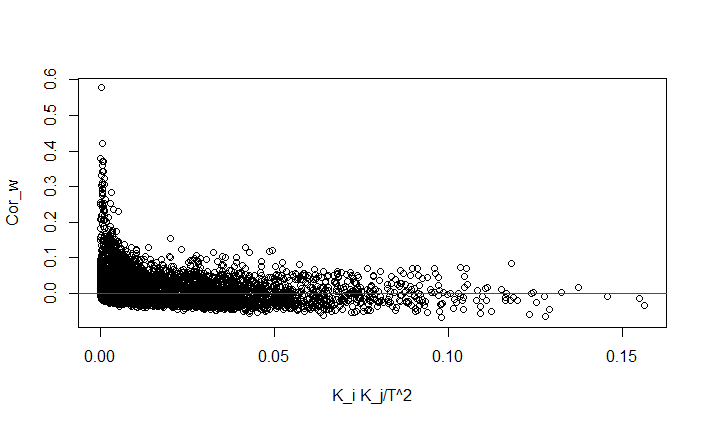}
    \caption[Weight-groups and odds-ratios of Finnish parliament synthetic data]{Simulation run with parameters similar to the Finnish parliament dataset, top row shows the ditribution of the row-sums of the data matrix and the segmented fit used to identify the weight-groups, mid row shows the plot of Newman's covariance (left) and weighted covariance (right) as a function of $K_i K_j/T^2$ and the bottom row shows the same plot of Newman's correlation and weighted one.}
    \label{fig:Finnish_synth}
\end{figure}

\begin{figure}
\center
	\includegraphics[width=0.49\textwidth]{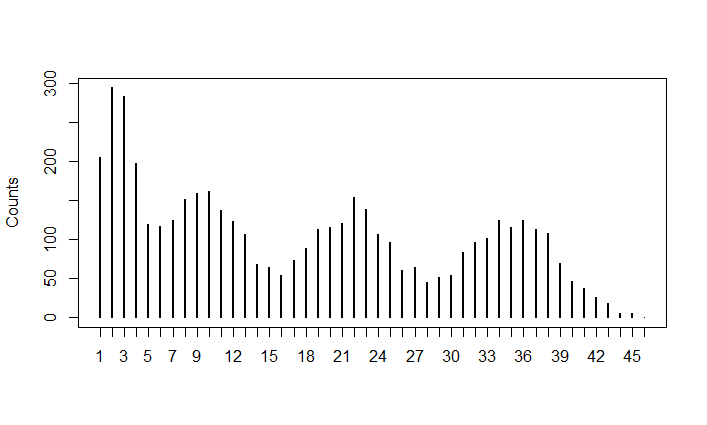}
	\includegraphics[width=0.49\textwidth]{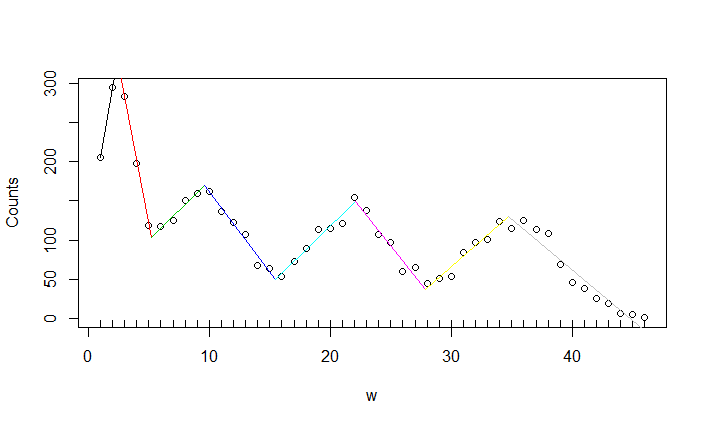}
	\includegraphics[width=0.49\textwidth]{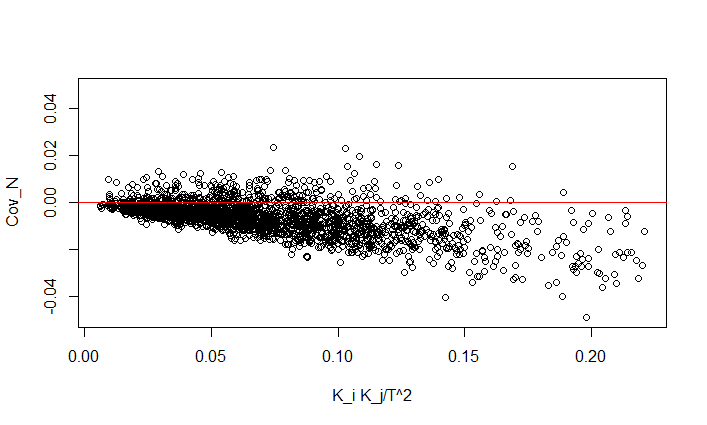}
	\includegraphics[width=0.49\textwidth]{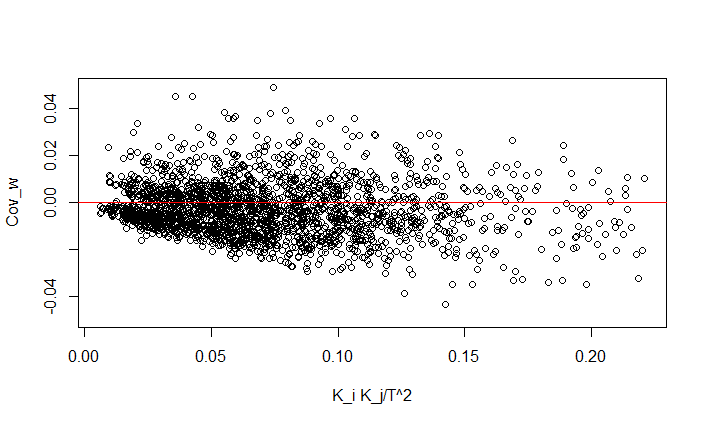}
	\includegraphics[width=0.49\textwidth]{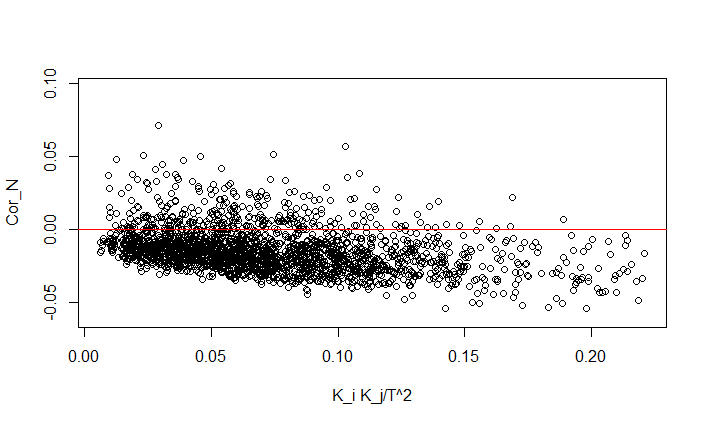}
	\includegraphics[width=0.49\textwidth]{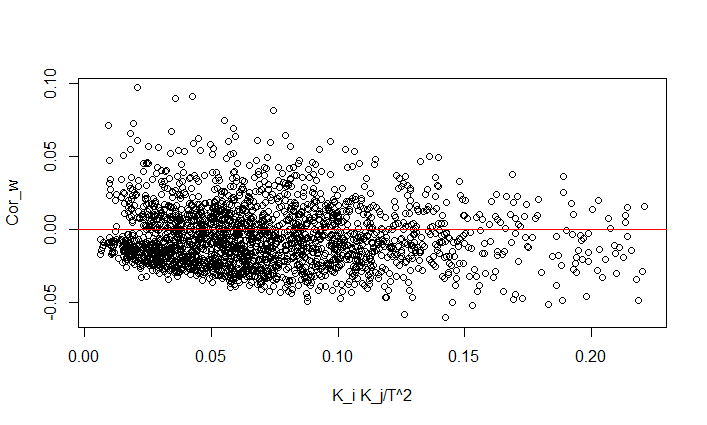}
    \caption[Weight-groups and odds-ratios of COGs synthetic data]{Simulation run with parameters similar to the COGs dataset, top row shows the ditribution of the row-sums of the data matrix and the segmented fit used to identify the weight-groups, mid row shows the plot of Newman's covariance (left) and weighted covariance (right) as a function of $K_i K_j/T^2$ and the bottom row shows the same plot of Newman's correlation and weighted one.}
    \label{fig:COG_synth}
\end{figure}

After exploring the spectrum of $K_i K_j/T^2 \in [0,1]$, we ran two simulations with initial parameters closer to our empirical datasets, by setting user's degrees exactly equal to those of each dataset, the weight-groups vector $\mathbf{m}=\{500,500,500,500\}$ for the Finnish parliament and $\mathbf{m}=\{1200,1200,1200,1200\}$ for the COGs dataset andthe odds-ratios vector $\mathbf{w}=\{0.05, 0.2, 0.5, 1\}$. The results are shown in Fig. \ref{fig:Finnish_synth} and Fig. \ref{fig:COG_synth}.

From all the simulations we ran, it's quite clear that the weighted estimators perform better than Newman's ones, which are still affected by a bias growing as user's degree increases. There are many other ways in which one can attempt to identify the weight-groups in empirical datasets when they are unknown a priori, but ours is quite simple and works well when the groups are not too superimposed.

\begin{figure}
\center
	\includegraphics[width=0.49\textwidth]{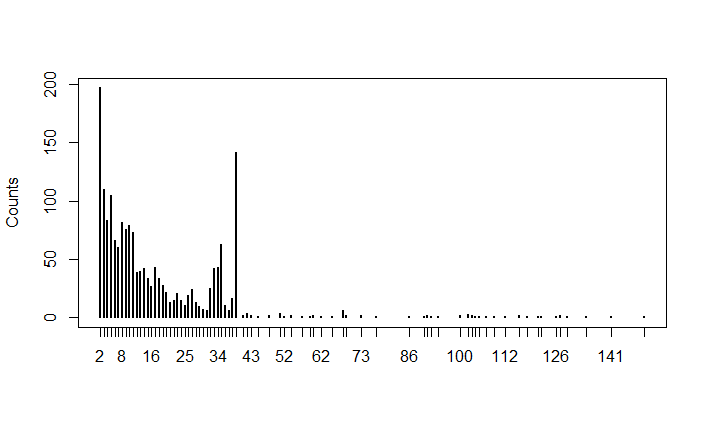}
	\includegraphics[width=0.49\textwidth]{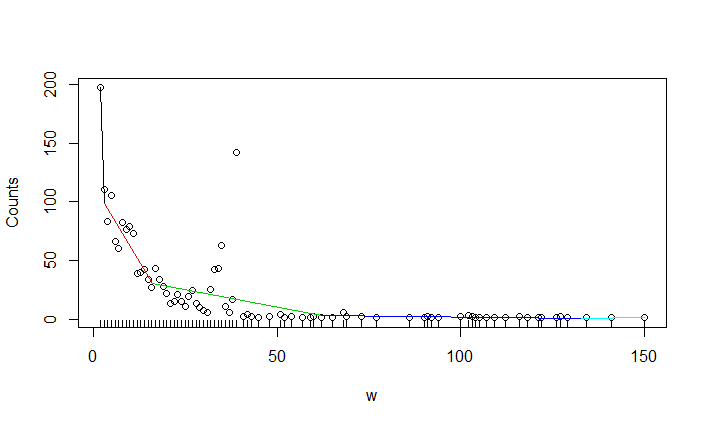}
	\includegraphics[width=0.49\textwidth]{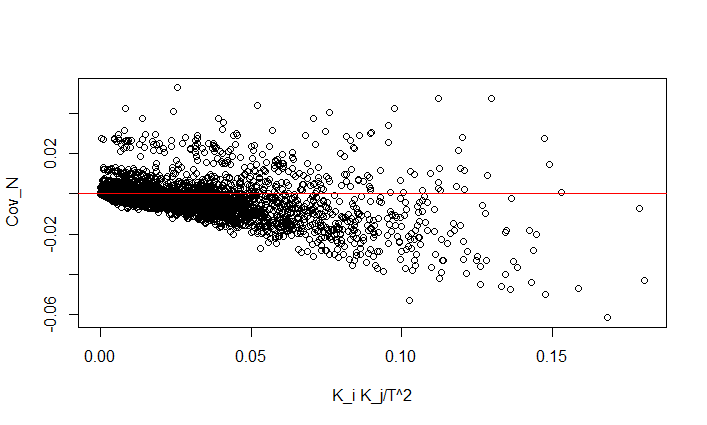}
	\includegraphics[width=0.49\textwidth]{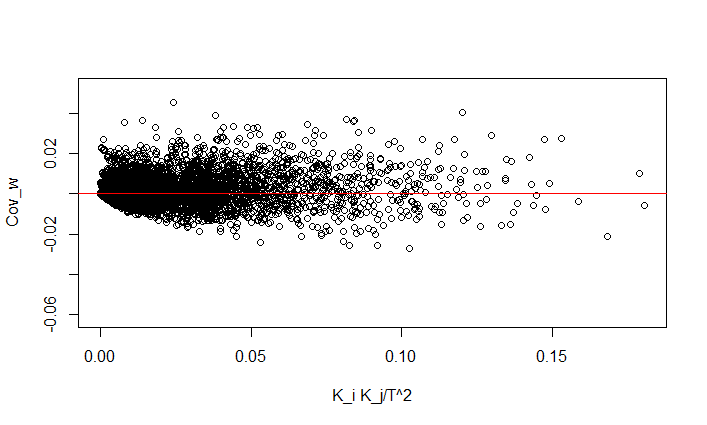}
	\includegraphics[width=0.49\textwidth]{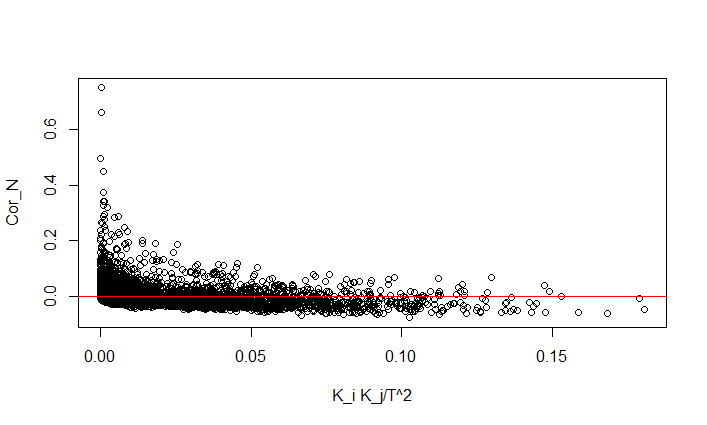}
	\includegraphics[width=0.49\textwidth]{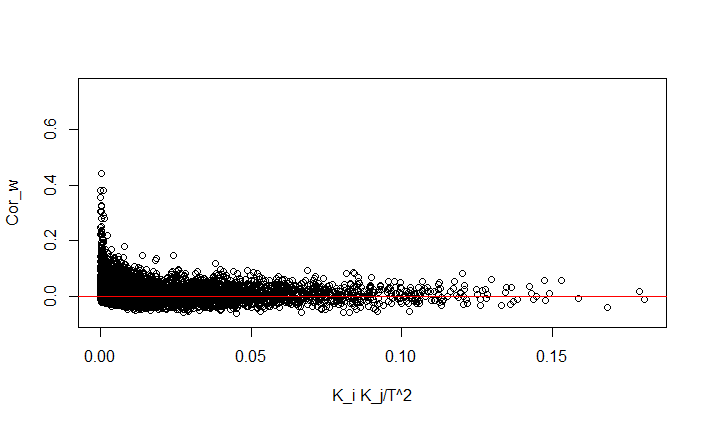}
    \caption[Weight-groups and odds-ratios of Finnish parliament rewired network]{Finnish parliament rewired matrix, top row shows the distribution of the row-sums of the real data rewired matrix and the segmented fit used to identify the weight-groups, mid row shows the plot of Newman's covariance (left) and weighted covariance (right) as a function of $K_i K_j/T^2$ and the bottom row shows the same plot of Newman's correlation and weighted one.}
    \label{fig:F_emp}
\end{figure}

\begin{figure}
\center
	\includegraphics[width=0.49\textwidth]{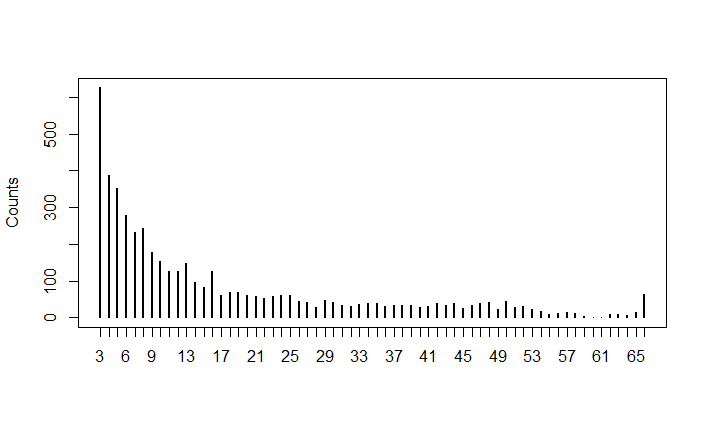}
	\includegraphics[width=0.49\textwidth]{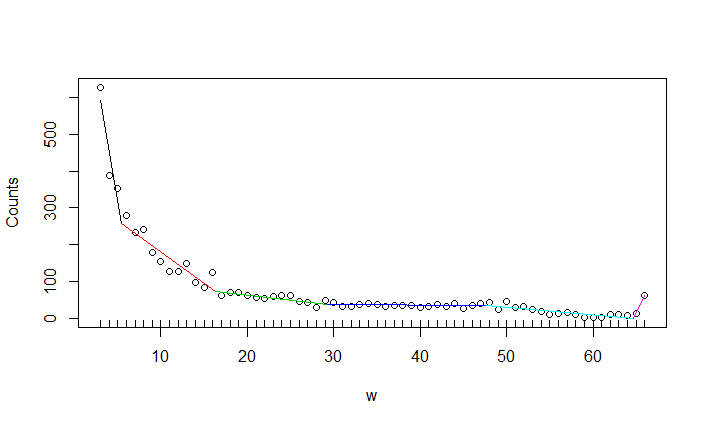}
	\includegraphics[width=0.49\textwidth]{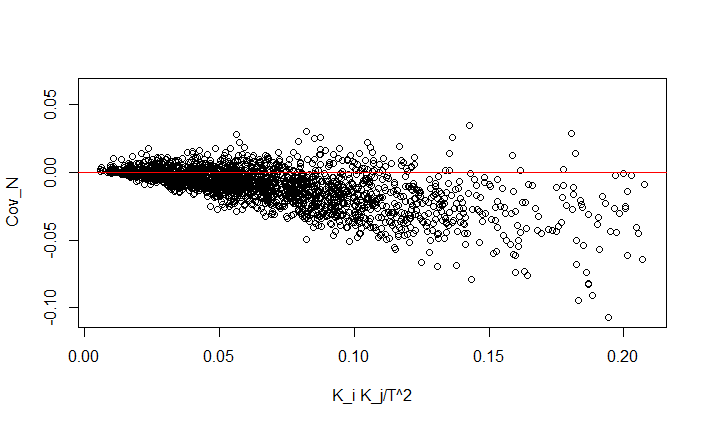}
	\includegraphics[width=0.49\textwidth]{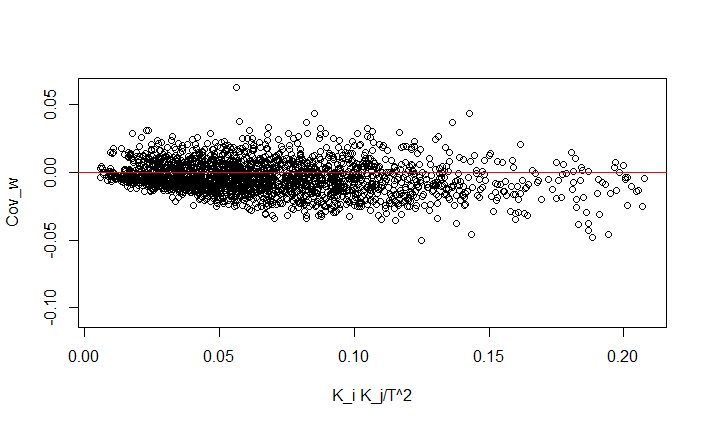}
	\includegraphics[width=0.49\textwidth]{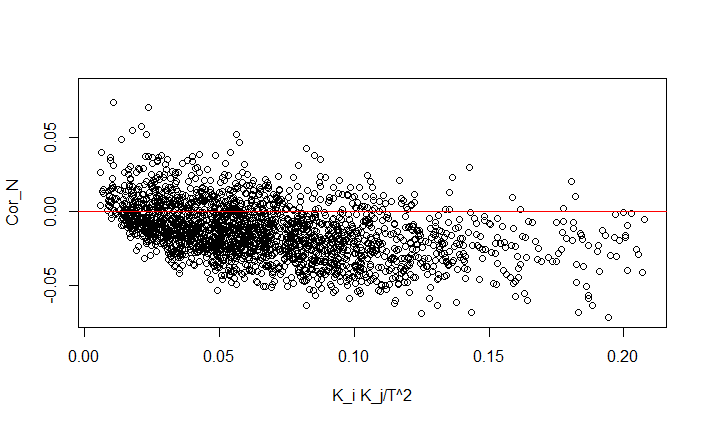}
	\includegraphics[width=0.49\textwidth]{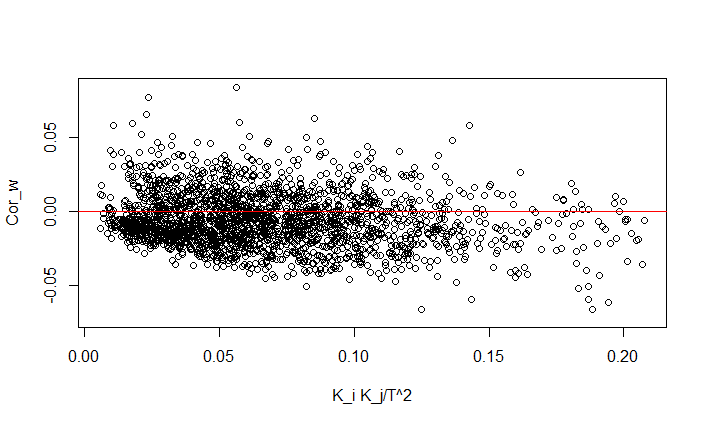}
    \caption[Weight-groups and odds-ratios of COGs rewired network]{COGs rewired matrix, top row shows the distribution of the row-sums of the real data rewired matrix and the segmented fit used to identify the weight-groups, mid row shows the plot of Newman's covariance (left) and weighted covariance (right) as a function of $K_i K_j/T^2$ and the bottom row shows the same plot of Newman's correlation and weighted one.}
    \label{fig:COG_emp}
\end{figure}

In Fig. \ref{fig:F_emp} and \ref{fig:COG_emp} we show the above described method to identify groups and relative odds-ratios for the rewired matrices of the Finnish parliament and COGs databases. The parameters we obtained from the algorithm are summarized in TABLE \ref{tab:parameters}.

\begin{table}
\begin{center}
\begin{tabular}{c | c c c c c c} 
\multicolumn{7}{c}{\textbf {Parameters obtained from the algorithm}}  \\ \hline
\multicolumn{7}{c}{\textit {Finnish parliament rewired network}}  \\ \hline
$psi$ & 3 & 16 & 62 & 133 & & \\
$m$ & 307 & 806 & 653 & 39 & 3 & \\
$w$ & 0.0096 & 0.0202 & 0.1040 & 0.7150 & 1 & \\
\hline
\multicolumn{7}{c}{\textit {COGs rewired network}}  \\ \hline
$psi$ & 5 & 16 & 29 & 48 & 65 & \\
$m$ & 1367 & 1792 & 718 & 669 & 250 & 77 \\
$w$ & 0.0040 & 0.0121 & 0.0378 & 0.1027 & 0.2558 & 1 \\
\end{tabular}
\caption[Summary of the parameters obtained from the algorithm]{\label{tab:parameters} Parameters obtained by running the segmentation algorithm on the $xy$ plot of the number of initiatives(COGs) with a given number of signatures(appearing in a given number of genomes), as a function of the corresponding number of signatures(number of genomes) and afterwards computing the odds-ratios. The break-points $psi$ retrieved by the segmentation algorithm are used to determine the weight-groups vector $m$, whose corrresponding odds-ratios vector $w$ is calculated according to Eq. (\ref{eq:oddsratio2}).}
\end{center}
\end{table}

From Fig. \ref{fig:F_emp} and \ref{fig:COG_emp} one can see how it is not always straightforward to single out the weight-groups in real systems. Nonetheless, even with a rough estimate, the weighted covariance and correlation coefficients are still closer to their expected values of zero than their Newman's
counterparts.

\section*{Conclusions}

We have provided evidence of biasing in the binary covariance and correlation estimators when applied to bipartite systems with a high degree of heterogeneity on both sides. Such a bias is already apparent in two systems, one social and the other biological, when looking at the correlation and covariance matrices of the randomly rewired network, which is supposed to be completely randomized, whereas both the correlation and covariance matrices turn out to be structured instead.

To explain the former structure and devise an unbiased estimator, we developed a simple theoretical model of the rewiring process, as a sampling without replacement from a biased urn, where the weights of the labeled marbles represent the degrees on the side of the network we're not projecting on. Therefore such a model is an approximation of the randomly rewired network, in the sense that afore mentioned degrees are preserved on average, according to the model, while they are exactly preserved in the rewired network. According to the biased urn model, two users randomly and independently pick a number of marbles equal to their degree, the underlying distribution being, therefore, the Wallenius non-central hypergeometric distribution. One can then calculate the expected value of random co-occurrence within each weight-category, that is the number of marbles with the same label randomly sampled by two users, by using the standard hypergeometric distribution. The model predicts a second order correction to the expected value of the covariance, which depends on both users degree and quadratically on the weight, when $w \simeq 1$.

The starting point to construct the unbiased estimator lies on the idea of getting rid of the weights by dividing the original users binary vectors, featuring a $1$ if there's a link between the user and the corresponding item on the other side of the network and a $0$ otherwise, by ad hoc weighting functions. The latter are chosen in such a way as to satisfy the requirement of zeroing the expected value of the covariance in the purely random case. By doing so, we automatically end up with a new estimator of covariance whose expectation value is zero under random rewiring, thus being unbiased. By using the same weighting functions used to estimate the covariance, the expected value of the correlation keeps showing a second order bias in $w$. However, such a bias is much smaller than the one in the unweighted estimator: it is $1/(K_i K_j)$ times the unweighted one, where $K_i$ and $K_j$ are the degrees of the considered users. Furthermore, from a more practical point of view, we've shown that such an improvement in the correlation estimator de facto zeroes the expected value of the correlation coefficient under rewiring as well, at least for a broad range of users' degrees. In both real-world examples analyzed in the paper.

Finally, the introduced covariance and correlation estimators perform better than the unweighted ones at grasping the clustered structure of the real bipartite networks considered in the paper. Specifically, they better capture aggregation by phyla in the COGs dataset and better discriminate between real and noise-induced clusters in the Finnish dataset of parliament initiatives.

In conclusion, our paper serves both as a warning to other researchers when using binary correlation and covariance to investigate bipartite systems with a high heterogeneity on both sides, and as a solution to the problem, in that we propose weighted estimators, which get rid of the bias problem. An R function that calculates weighted correlation and covariance of a bipartite system is provided in the supplementary information to the paper.  

The biased urn model we used to introduce weighted estimators of correlation and covariance could also be used to associate a level of statistical significance with co-occurrence in bipartite system, which might have implications on the construction of statistically validated networks. However, that is an ongoing research and out of the scope of the present paper.

\end{document}